\documentclass[journal=jpcbfk,manuscript=article]{achemso}
\usepackage{amssymb,amsmath}
\usepackage[normalem]{ulem}
\usepackage{bm}
\usepackage{color}
\newcommand{\angstrom}{\text{\normalfont\AA}}
\usepackage{hyperref}
\usepackage[dvipsnames]{xcolor}
\usepackage{setspace}
\usepackage{amsmath}
\usepackage{amsfonts}
\usepackage{listings}
\usepackage{graphicx}
\usepackage{hyperref}
\usepackage{makecell}

\mciteErrorOnUnknownfalse

\SectionNumbersOn

\newcommand*{\citen}[1]{%
  \begingroup
    \romannumeral-`\x 
    \setcitestyle{numbers}%
    \cite{#1}%
  \endgroup   
}

\author{Irem Altan} 
\affiliation{Department of Chemistry, Duke University, Durham, NC 27708, USA}
\author{Jennifer McManus}
\affiliation{Department of Chemistry, Maynooth University, Maynooth, Ireland}
\author{Patrick Charbonneau}
\affiliation{Department of Chemistry, Duke University, Durham, NC 27708, USA}
\alsoaffiliation{Department of Physics, Duke University, Durham, NC 27708, USA}
\email{irem.altan@duke.edu}
\date{\today}

\title{Using schematic models to understand the microscopic basis for inverted solubility in $\gamma$D-crystallin}

\begin{document}
\begin{abstract}
Inverted solubility--a crystal melting upon cooling--is observed in a handful of proteins, such as carbomonoxy hemoglobin and $\gamma$D-crystallin. In human $\gamma$D-crystallin, the phenomenon is associated with the mutation of the 23$^\mathrm{rd}$ residue, a proline, to a threonine, serine or valine. One proposed microscopic mechanism for this effect entails an increase in hydrophobicity upon mutagenesis. Recent crystal structures of a double mutant that includes the P23T mutation allows for a more careful investigation of this proposal. Here, we first measure the surface hydrophobicity of various mutant structures of this protein and determine that it does not discernibly increase upon the mutating the 23$^\mathrm{rd}$ residue. We then investigate the solubility inversion regime with a schematic patchy particle model that includes one of three models for temperature-dependent patch energies: two of the hydrophobic effect, and a more generic description. We conclude that while solubility inversion due to the hydrophobic effect may be possible, microscopic evidence to support it in $\gamma$D-crystallin is weak. More generally, we find that solubility inversion requires a fine balance between patch strengths and the temperature-dependent contribution, which may explain why inverted solubility is not commonly observed in proteins. In any event, we also find that the temperature-dependent interaction has only a negligible impact on the critical properties of the $\gamma$D-crystallin, in line with previous experimental observations.
\end{abstract}

\maketitle

\section{Introduction}
\label{sec:introduction}
Proteins can self-organize into a rich variety of superstructures~\cite{Mcmanus:2016}, such as crystals~\cite{Fusco:2016}, virus capsids~\cite{Hagan:2014}, disease-forming aggregates~\cite{Saric:2014}, and biomaterials~\cite{Suzuki:2016}. A key challenge is to understand how microscopic features of solvated proteins can give rise to such complex phase diagrams, and eventually to design systems that reliably assemble~\cite{Pieters:2016,Glotzer:2007,Huang:2016,Suzuki:2016,Salgado:2010,Koehler:2015}.
In this context, coarse-grained models are especially valuable, because they help both pinpoint and abstract the microscopic features that can reproduce the experimentally observed behavior. (Since simulating protein self-assembly typically requires hundreds to thousands of protein copies, which are themselves comprised of thousands of atoms, such models are also a computational necessity~\cite{Altan:2019,Vega2008,Romano:2010}.)
For example, even simple models of short-ranged~\cite{tenWolde:1997,Romano:2009}, anisotropic pair interactions can largely explain the phase behavior of globular proteins~\cite{Lomakin:1999,Bianchi:2011,Mcmanus:2016}. 
Understanding the assembly of some systems, however, requires coarse-grained models with additional features, such as shape anisotropy for viral capsid and amyloid forming proteins~\cite{Hagan:2014,Mcmanus:2016}. 
Capturing certain features of protein crystallization, which is key to protein structure determination by diffraction methods~\cite{McPherson:2004}, also requires enhanced patchy particle models~\cite{Fusco:2016}.

Systems that exhibit atypical phase behaviors are essential test of our understanding of the physico-chemical processes that underlie protein assembly. One such phenomenon is inverted crystal solubility, i.e., the decrease of solubility with increasing temperature. This phenomenon is observed in a handful of proteins, such as some single mutants of $\gamma$D-crystallin~\cite{Pande:2009,Pande:2010}, and the wild type carbomonoxy-hemoglobin C~\cite{Vekilov:2002}. (The temperature invariant solubility of apoferritin is a limit case~\cite{Petsev:2001}.) Thermodynamically, inverted solubility suggests that as temperature increases, the Gibbs free energy of crystallization decreases, and hence that the crystal becomes increasingly more stable than the fluid. The phenomenon is often attributed to a large and positive entropy gain upon crystallization. Crystal formation is then possible despite the enthalpy of crystallization being non-negative~\cite{Vekilov:2002,Shiryayev:2005,Wentzel:2008}.  Because the solute contribution to the change in entropy is typically negative, the solvent contribution is considered to be the key microscopic determinant~\cite{Wentzel:2008,Shiryayev:2005,Vekilov:2002}. 

The association of inverted solubility in proteins with the hydrophobic effect comes from our understanding of the aqueous solvation of hydrocarbons, which presents an analogous solubility regime~\cite{Moelbert:2003}. 
A minimal model for this effect was proposed by Lee and Graziano~\cite{Lee:1996}, who refined Muller's two state description of water as having either active or broken hydrogen bonds~\cite{Muller:1990}. 
The final model formulation considers water as being in one of four states: disordered shell (ds), ordered shell (os), disordered bulk (db), and ordered bulk (ob). 
Shiryayev et al.~then used this model to estimate the phase diagram of model globular proteins with isotropic interactions assumed to be driven exclusively by hydrophobic interactions~\cite{Shiryayev:2005}.  Although the resulting phase behavior does present an inverted solubility regime, 
it is unclear whether it persists for more realistic protein models, with directional interactions and a complex surface mosaic of hydrophilic and hydrophobic interactions with the solvent. 
In other words, while the hydrophobic scenario for solubility inversion in proteins is thermodynamically sound, microscopic evidence for it remains limited. The generality of the underlying physical arguments is also seemingly incompatible with the relatively rare occurrence of inverted solubility in experiments. 

Here, we examine this microscopic scenario in the context of a double mutant (R36S+P23T) of the human $\gamma$D-crystallin, which has been shown to form two competing crystals: a normal solubility structure (DBN, PDB~\cite{Berman:2000} ID: 6ETA) and an inverted solubility structure (DBI, PDB ID: 6ETC)~\cite{James:2015,Khan:2019}. The solubility inversion is here most likely associated with the mutation in the 23$^\mathrm{rd}$ residue because the single mutants P23T, P23S, and P23V also exhibit inverted solubility~\cite{Mcmanus:2007}. 
In earlier work, we have parameterized a patchy model for this system and have obtained a solubility inversion regime by completely deactivating the patch containing the 23$^\mathrm{rd}$ residue at low temperatures~\cite{Khan:2019}. 
Interestingly, the DBI crystal does not present any obvious structural feature, other than the formation of a hydrogen bond through the 23$^\mathrm{rd}$ residue.
Here, we test three different temperature-dependent interaction potentials: the generic model we previously considered, and two that explicitly model the hydrophobic scenario. We use these models to test the hydrophobic scenario as well as the robustness of the inverted solubility regime with respect to model parameters. We thus attempt to elucidate why inverted solubility is not more commonly observed. We further explore the relationship between the liquid-liquid critical point and the solubility curve which has been studied experimentally for some of these systems~\cite{Mcmanus:2007}. The plan for the rest of this paper is as follows. 
We first consider whether an increase in surface hydrophobicity can be discerned upon introducing the solubility inverting mutations (Sec.~\ref{sec:surfHydro}).
We then introduce a patchy protein model for these proteins (Sec.~\ref{subsubsec:patchyModel}) along with the different temperature-dependent patch models (Sec.~\ref{subsec:invSolubility Models}), and the methods used to determine solubility lines (Sec.~\ref{subsubsec:phaseDiagramMethod}). 
Sections \ref{subsub:nw} and \ref{subsubsec:tempDeacResults} provide a detailed analysis of these patchy models, and we conclude with proposals for further discerning experiments in Section~\ref{sec:conclusions}.

\section{Surface Hydrophobicity}
\label{sec:surfHydro}

As a first consideration of the reasonableness of the hydrophobicity scenario, we first evaluate the surface hydrophobicity of various human $\gamma$D-crystallin crystal structures. Were the P23T mutation to consistently increase surface hydrophobicity, it would serve as strong evidence for the decrease in protein solubility upon mutagenesis to be driven by the hydrophobic effect.
By studying the relative binding propensity of two dyes known to bind hydrophobic surfaces, Pande et al.~have inferred that P23T, P23S, and P23V mutants of human $\gamma$D-crystallin do present a higher surface hydrophobicity than the wild type (WT) protein.~\cite{Pande:2010}
In order to test the robustness of this interpretation, we here consider different scales that quantify hydrophobicity at the amino acid level. More specifically, we compute an average of hydrophobicity indices of solvent-exposed residues~\cite{PyMol} 
weighted by their solvent accessible surface area (SASA)~\cite{Fusco:2014b}, for five different scales: grand average of hydropathy (GRAVY)~\cite{Kyte:1982}, as well as the scales of Wimley and White (\textbf{ww})~\cite{Wimley:1996}, Hessa et al. (\textbf{hh})~\cite{Hessa:2005}, Moon and Fleming (\textbf{mf})~\cite{Moon:2011}, and Zhao and London (also known as transmembrane tendency, \textbf{tt})~\cite{Zhao:2006}. Each of these scales assigns a hydrophobicity index to each residue type; all but \textbf{hh} and \textbf{mf} assign positive values to hydrophobic residues.

We compute hydrophobicity for three sets, $S$, of amino acids: (i) the entire protein surface, (ii) the surface of its N-terminus, i.e., the first 82 residues (including the solubility inverting 23$^\mathrm{rd}$ residue), and (iii) the surface residues in the DBI contact that includes the 23$^\mathrm{rd}$ residue (Patch 4 as per Sec.~\ref{sec:solFromPatchy})~\cite{Khan:2019}.
The hydrophobicity, $H_\zeta$, for a given scale $\zeta$ is then obtained as

\begin{gather}
	H_\zeta = \frac{\sum\limits_{i\in S}f_\zeta(i)A(i)}{\sum\limits_{i\in S} A(i)},
\end{gather}
where $f_\zeta(i)$ is the hydrophobicity index for residue $i$, and $A(i)$ is its SASA. 
We specifically consider: WT (PDB ID: 1HK0~\cite{Basak:2003}), the P23T single mutant (PDB ID: 4JGF~\cite{Ji:2013}), the R36S single mutant (PDB ID: 2G98~\cite{Kmoch:2000}), the R58H single mutant (PDB ID: 1H4A~\cite{Basak:2003}), DBI (PDB ID: 6ETC~\cite{Khan:2019}), and DBN (PDB ID: 6ETA~\cite{Khan:2019}). Of these, only WT, R36S, and R58H do not have a mutation at the 23$^\mathrm{rd}$ residue. 
Note that  missing residues are completed using Modeller~\cite{sali:1993} within Chimera~\cite{Pettersen:2004}, and all crystal water molecules are removed. 
In order to estimate the error on these measured hydrophobicities, 100 structures per crystal are created by perturbing each particle coordinate by a random number selected from a Gaussian distribution with standard deviation corresponding to the coordinate error specified in the PDB file. 
Two assumptions are made in estimating these error bars. First, the coordinate error reported in the PDB entry is assumed to be distributed uniformly and isotropically across all protein atoms. In reality, certain domains or residues in proteins are more mobile and thus harder to resolve by X-ray diffraction than others, but residue-level information is not available. This assumption thus overestimates the error in more localized parts of the protein and underestimates the error in more mobile parts. Second, the refined structures do not precisely capture the actual protein structure, as suggested by $R_\mathrm{free}$ values ranging from 0.174 to as high as 0.284, hence possibly creating artificial hydrophobicity differences between different mutants, or, conversely, underestimating them.

\begin{figure}
	\includegraphics[width=\textwidth]{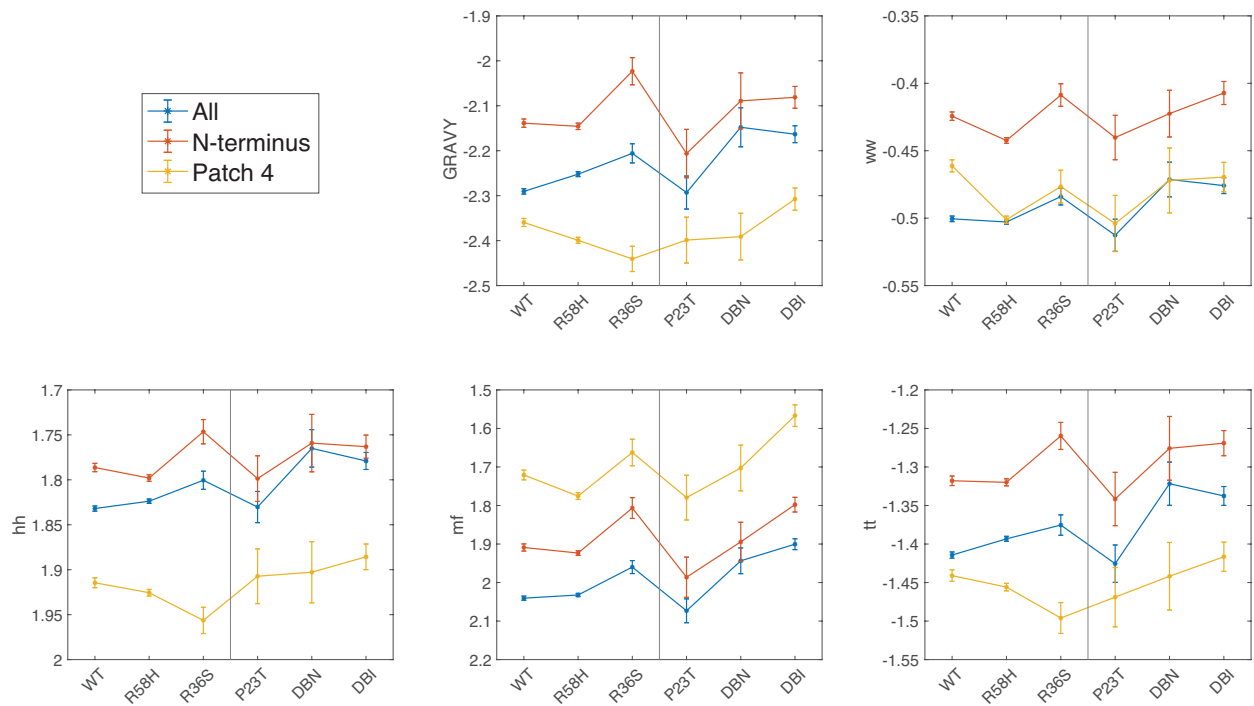}
	\caption{Hydrophobicity estimates for different crystal structures of single and double mutants of human $\gamma$D-crystallin. Proteins to the left of the black vertical line exhibit normal solubility, and those to the left exhibit inverted solubility. The error bars denote $95\%$ confidence intervals. Lines connecting the points are but a guide to the eye. Note the flipped scales for the  \textbf{hh} and \textbf{mf} scales,  in which lower values denote higher hydrophobicity, by contrast to the other scales. Structures with the P23T mutation do not systematically present a higher hydrophobicity, which is inconsistent with the hydrophobicity scenario.}
	\label{fig:hydrophobicities}
\end{figure}

The resulting hydrophobicity estimates are shown in Fig.~\ref{fig:hydrophobicities}.
We first compare the DBN and DBI structures, which are obtained from the same double mutant, R36S+P23T, and which are structurally very similar~\cite{Khan:2019}. As expected, nearly all measurements for DBN and DBI overlap within their $95\%$ confidence intervals. The only exception is the hydrophobicity of Patch 4 measured by the \textbf{mf} scale. This could be because \textbf{mf} uniquely classifies prolines as hydrophobic. 
This discrepancy could then amplify the minute difference in surface exposure of Patch 4 prolines between DBI and DBN.

Overall, the N-terminus is the most hydrophobic region in nearly all scales and for all structures. However, other observations are not consistent across scales.  A number of nonmonotonicities can indeed be observed across different hydrophobicity scales. For instance, Patch 4 is more hydrophobic in DBI than in WT in the GRAVY, \textbf{hh}, and \textbf{mf} scales, but the \textbf{ww} and \textbf{tt} scales present no discernible difference. Similarly, Patch 4  is more hydrophobic in R36S than WT for the \textbf{mf} scale, but the reverse is true for \textbf{hh}.
These discrepancies reflect the different ordering of residues in the various hydrophobicity scales.
For instance, GRAVY, which is calculated from experimental measurements of transfer free energies from water to water vapor, tends to assign aromatic side chains lower hydrophobicities than the other four scales, which instead consider the tendency of residues to transfer from water to within a lipid bilayer, a measurement prone to more experimental uncertainty~\cite{Wimley:1996}.

Interestingly,  the N-terminus of the P23T mutant is the least hydrophobic structure for the GRAVY and \textbf{mf} scales. This trend, however, disappears when only Patch 4 residues are considered.
Patch 4, which controls solubility inversion, is actually \emph{less} hydrophobic than the overall N-terminus or the entire protein, except on the \textbf{mf} scale. Only for this last scale is Patch 4 clearly more hydrophobic. A similar inconsistency exists for Patch 4 of DBI, which is more hydrophobic than the other proteins for GRAVY and \textbf{mf}, but for these two scales P23T and DBN are not discernibly more hydrophobic than the structures without the mutation in the 23$^\mathrm{rd}$ residue. 

In summary, in none of the hydrophobicity scales do the structures with the (solubility-inverting) P23T mutation have a statistically and consistently higher hydrophobicity than those without. 
P23T mutations even result in \emph{lower} hydrophobicity estimates on some scales. 
While our measurements are subject to errors from the crystal structure accuracy, as well as the imperfections of the hydrophobicity scales themselves, a microscopic change to the protein surface that could underlie inversion of solubility remains elusive from this viewpoint. 
The question that then arises is whether other measurements could be more revealing. Because hydrophobicity scales are but an indirect measure of protein-water interactions (and thus protein-protein interactions), more direct measurement could be microscopically more revealing. 
In that context, we next consider the hydrophobicity scenario from a thermodynamic perspective.

\section{Solubility Lines from Patchy Models}
\label{sec:solFromPatchy}
Because a clear enhancement of hydrophobicity cannot be detected directly in mutants with inverted solubility, we next consider the thermodynamics of patchy models that incorporate various temperature-dependent patch energies. A schematic model of the double mutant of human $\gamma$D-crystallin was previously studied in Ref.~\citen{Khan:2019}, but it is here modified to consider the hydrophobic scenario and then perturbed to evaluate the robustness of its inverted solubility regime.

\subsection{Patchy Model}
\label{subsubsec:patchyModel} 
\begin{figure}
	\includegraphics[width=0.4\textwidth]{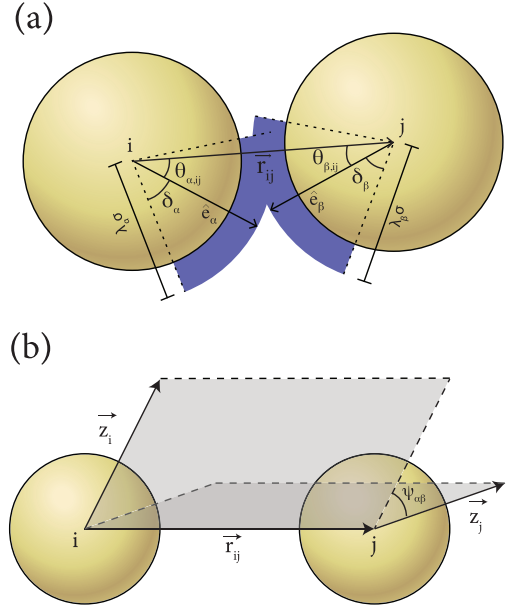}
	\caption{For two patches to interact, the relative particle orientation should satisfy the following. (a) The angle between the vector joining particles $i$ and $j$, $\mathbf{r}_{ij}$, and the patch vectors $\mathbf{\hat{e}}_\alpha$ and $\mathbf{\hat{e}}_\beta$ should be less than $\delta_\alpha$ and $\delta_\beta$, respectivel. (b) The dihedral angle between two particles, which is defined as the angle between two planes defined by the vectors $(\mathbf{z}_i, \mathbf{r}_{ij})$ and $(\mathbf{z}_j, -\mathbf{r}_{ij})$, should be within the range $\varphi_{ab}\pm\Delta\varphi_{ab}$. The reference vector $\mathbf{z}$ is chosen such that its orientation relative to the patches is identical for all particles.}
	\label{fig:orientationTorsion}
\end{figure}

The schematic model consists of hard particles with attractive patches 

\begin{gather}
	u(r_{ij},\Omega_i, \Omega_j) = u_\mathrm{HS}(r_{ij}) + \sum\limits_{a,b}^n u_{ab}(r_{ij},\Omega_i, \Omega_j),
\end{gather}
where $r_{ij}$ is the distance between particles $i$ and $j$, $\Omega$ denotes the particle orientation, and $u_\mathrm{HS}(r_{ij})$ is the hard sphere potential for particles of diameter $\sigma$. The sum runs over all patch pairs, with $n$ the total number of patches. The second contribution, $u_{ab}$, is further broken down into radial and angular parts

\begin{gather}
	u_{ab}= v_{ab}(r_{ij})f_{ab}(\Omega_i,\Omega_j).
\end{gather}
The radial part, $v_{ab}$, is a square-well interaction

\begin{gather}
	v_{ab}(r_{ij}) =\left\{ \begin{array}{ll}
		-\varepsilon_{ab}(T), & \sigma < r_{ij} < \lambda_a+\lambda_b \\
		0, &\text{otherwise}
	\end{array} \right. ,
\end{gather}
with interaction ranges $\lambda_a$ and $\lambda_b$ of patches $a$ and $b$, respectively, and with either constant or temperature-dependent patch energy $-\varepsilon_{ab}(T)$. The orientational part

\begin{gather}
	f_{ab}=\left\{ \begin{array}{ll}
		1, & \theta_{a,ij} \leq \delta_a\;\mathrm{and}\; \theta_{b,ij} \leq \delta_b \\
		0, &\text{otherwise}
	\end{array} \right.
	\times
	\left\{ \begin{array}{ll}
		1, & \psi_{ij}\in[\varphi_{ab}-\Delta\varphi_{ab}, \varphi_{ab}+\Delta\varphi_{ab}] \\
		0, &\text{otherwise}
	\end{array} \right.
\end{gather}
contains two contributions. The first ensures that the relative particle orientation enables them to interact with $\delta_a$ and $\delta_b$ the angular width for patches $a$ and $b$, respectively (Fig.~\ref{fig:orientationTorsion}a). The second limits the range   $\varphi_{ab}\pm \Delta\varphi_{ab}$ of dihedral angles $\psi_{ij}$ allowed for each pair  (Fig.~\ref{fig:orientationTorsion}b), with $\theta_{a,ij}$  the angle between the vector defining the location of patch $a$ and the vector that connects the centers of particles $i$ and $j$, and $\theta_{b,ij}$ similarly for patch $b$. 

This model is parameterized such that each patch corresponds to a crystal contact in either the DBI or DBN crystal structure. This choice assumes that these surface patches are most chemically relevant for crystal formation, which is reasonable for such a small protein and is consistent with earlier studies of protein crystallization~\cite{Fusco:2014}. We then obtain five patches for DBI, labeled with Arabic numerals, and five patches for DBN, labeled with Roman numerals. Because Patch 4 of DBI contains the 23$^\mathrm{rd}$ residue, which is associated with the inverted solubility regime, this patch is taken to be temperature dependent (see Sec.~\ref{subsec:invSolubility Models}); other patches have a constant energy. Patch energies and interaction ranges were previously extracted from all-atom molecular dynamics simulations~\cite{Berendsen:1995}, using umbrella sampling~\cite{Kastner:2011}. The resulting effective single-component system model thus coarse-grains the role of solvent and ions in the crystallization cocktail.

(The resulting patchy particle model is sketched in Fig.~\ref{fig:patchSketch}, and the geometry details are given in the supplementary information.) In what follows, unless otherwise specified, energies are reported in units of $k_\mathrm{B}T_\mathrm{ref}$, where $T_\mathrm{ref}=277$K is the temperature at which DBN was crystallized experimentally, and distances are reported in units of the particle diameter $\sigma$, which here is taken to be 2.54 nm.

It is important to highlight that this protocol presents a number of limitations, including inaccuracies of the protein force field~\cite{Wickstrom:2009} and of the water model~\cite{Altan:2018}, inefficient sampling, as well as the crudeness of representing potentials of mean force as square well interactions and proteins as spheres. In addition, determining the potential of mean force for each crystal contact is a computationally challenging task, and the 10 ns sampling of each umbrella window likely incompletely explores some of the protein conformational changes, such as loop motion~\cite{Henzler:2007}. 
On the whole, this approach likely yields estimates of protein-protein interactions that are at best within 10 to 50\% of the association free energy. If sufficiently robust, the resulting phase diagram should therefore be qualitatively, although not quantitatively captured.

\begin{figure}
	\includegraphics[width=0.5\textwidth]{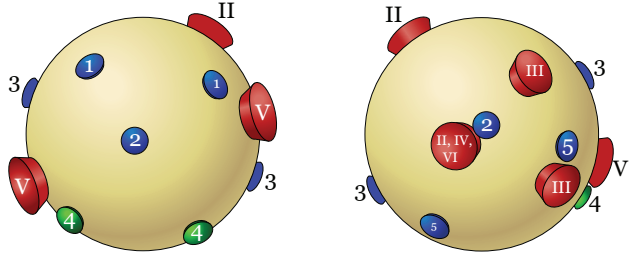}
	\caption{Front and back views of the patchy particle model. Blue and green patches are derived from DBI, and red patches from DBN. Patch 4 (green) contains the mutation associated with solubility inversion.}
	\label{fig:patchSketch}
\end{figure}

\subsection{Inverted Solubility Models}
\label{subsec:invSolubility Models}

In order to represent the microscopic origin of the inverted solubility, we consider three models for the temperature-dependence of Patch 4: the MLG model, the Wentzel-Gunton model, and the temperature-(de)activated patchy model. Note that the parameters explicitly defined in these models are provided here, while the free parameters are left for discussion in Sec.~\ref{subsub:nw} 

\emph{MLG model--} 
In this model, each of four water states is assigned an energy, $E$, and a (relative) degeneracy, $q$. Degeneracies are ordered $q_\mathrm{ds}>q_\mathrm{db}>q_\mathrm{ob}>q_\mathrm{os}$. The last inequality follows from the hydration shell allowing the formation of only hydrogen bonds between water molecules and not to the hydrophobic solute. The higher degeneracy of the disordered shell compared to the disordered bulk follows from the additional orientational constraints in the former compared to the latter. Because only relative information about the degeneracies is needed, the estimates of Ref.~\citen{Silverstein:1999} here suffice: $q_\mathrm{ob}=1.5$, $q_\mathrm{db}=30$, $q_\mathrm{os}=1$, and $q_\mathrm{ds}=48$. The energies are ordered $E_\mathrm{ds}>E_\mathrm{db}>E_\mathrm{ob}>E_\mathrm{os}$. The ordered shell is expected to have a lower energy than the ordered bulk state, because hydrogen bonds that form via tangentially oriented water molecules tend to be stronger than radially oriented ones. The disordered shell is expected to have a higher energy than the disordered bulk because replacing the solute with water molecules increases slightly the number of hydrogen bonds.
Because energy values used by Ref.~\citen{Moelbert:2003} are on an arbitrary  scale, which is incompatible with the specific energy scale of our patchy model, we use instead $E$ and $q$ reported by Silverstein et al.~for the Mercedes-Benz model of water~\cite{Silverstein:1998,Silverstein:1999}. Posing that the energy of the ordered bulk is about one hydrogen bond, $E_\mathrm{ob}=-5.82$ $k_\mathrm{B}T_\mathrm{ref}$~\cite{Feyereisen:1996},  
the other three states have: $E_\mathrm{db}=-1.69$ $k_\mathrm{B}T_\mathrm{ref}$, $E_\mathrm{os}=-5.90$ $k_\mathrm{B}T_\mathrm{ref}$, and $E_\mathrm{ds}=-0.56$ $k_\mathrm{B}T_\mathrm{ref}$. The energy and entropy per water molecule in the shell and are then given as~\cite{Shiryayev:2005}

\begin{gather}
E_s = \frac{E_\mathrm{os}+E_\mathrm{ds}e^{-\beta(E_\mathrm{ds}-E_\mathrm{os})}}{1+e^{-\beta(E_\mathrm{ds}-E_\mathrm{os})}}\\
E_b = \frac{E_\mathrm{ob}+E_\mathrm{db}e^{-\beta(E_\mathrm{db}-E_\mathrm{ob})}}{1+e^{-\beta(E_\mathrm{db}-E_\mathrm{ob})}}\\
\end{gather}
and
\begin{gather}
	S_s/k_B = \log\Big(\frac{q_\mathrm{os}+q_\mathrm{ds}e^{-\beta(E_\mathrm{ds}-E_\mathrm{os})}}{1+e^{-\beta(E_\mathrm{ds}-E_\mathrm{os})}}\Big)\\
	S_b/k_B = \log\Big(\frac{q_\mathrm{ob}+q_\mathrm{db}e^{-\beta(E_\mathrm{db}-E_\mathrm{ob})}}{1+e^{-\beta(E_\mathrm{db}-E_\mathrm{ob})}}\Big).
\end{gather}
The differences in energy and entropy per water molecule in the solvation shell of the protein and in the bulk are then simply $\varepsilon_w=E_s-E_b$ and $\Delta s_w = S_s - S_b$, respectively.

With this formulation the energy of Patch 4 is given by 

\begin{gather}
	\varepsilon_4^\prime = \varepsilon_4 + n_w\Delta\varepsilon(\beta),
\end{gather}
where we have defined $\Delta\varepsilon(\beta) =2(\varepsilon_w - \Delta s_w/\beta)$, and $n_w$ is the number of water molecules in the solvation shell around contact $i$.
Note that because patch parameters are measured at $\beta_\mathrm{ref}=1$, parameters need to be tuned such that $\varepsilon_4^\prime(\beta=1)=\varepsilon_4$, and hence
$\varepsilon_4^\prime = (\varepsilon_4-\Delta\varepsilon(1)n_w) + \Delta\varepsilon(\beta)n_w$. Note also that the temperature scale for the MLG model cannot be changed arbitrarily by changing $\beta_\mathrm{ref}$, because its parameters already set the range of temperatures within which the hydrophobic effect changes the free energy of crystallization.

\emph{Wentzel-Gunton Model--}  Wentzel and Gunton proposed a simplified version of the MLG model in order to consider the phase behavior of anisotropic particles using Wertheim's theory~\cite{Wertheim:1984,Wertheim:1984b,Sear:1999,Wentzel:2008}. This simple model assigns a linear temperature dependence for the patch energies

\begin{gather}
	-\varepsilon_4^\prime = -\varepsilon_4 - 2 \varepsilon_w + \frac{2}{\beta}\Delta s_w,
	\label{eq:gunton}
\end{gather}
where $-\varepsilon_w$ and $-\Delta s_w$ are free parameters that account for the change in energy and in entropy, respectively, due to the displacement of water upon contact association.  Patch energies should equal those of the original model at $\beta=\beta_\mathrm{ref}$, at which the model is parameterized. By contrast to the MLG model, this choice here suffices to set the overall temperature scale, because $\varepsilon_w$ and $\Delta s_w$ are arbitrary. Fixing $\varepsilon_w$, such that $\varepsilon_4^\prime(\beta=\beta_\mathrm{ref}) = \varepsilon_4$, thus results in $\varepsilon_4^\prime = \varepsilon_4 + 2\Delta s_w (\frac{1}{\beta_\mathrm{ref}}-\frac{1}{\beta})$.

\emph{Temperature-(de)activated Patchy Model--} de Las Heras and de Gama~\cite{de:2016} proposed a model for patch (de)activation with temperature inspired by DNA-grafted colloids, which lose their attractive patches above the DNA melting temperature~\cite{Geerts:2010}, 
Although this model does not correspond to a specific microscopic scenario in proteins,
it can nevertheless be construed as a simple and elegant way to describe patch (de)activation.
The temperature dependence of the interaction is then

\begin{gather}
	\varepsilon_4^\prime(T) = \frac{\varepsilon_4}{2}\Bigg[ 1 + \tanh \Bigg( \frac{T-T_a}{\tau} \Bigg) \Bigg],
\end{gather}
where $T_a$ is the deactivation temperature, $\tau$ controls the sharpness of that deactivation. For this model, Patch 4 is deactivated below $T_a$. 

\subsection{Crystal Solubility Determination}
\label{subsubsec:phaseDiagramMethod}

Solubility lines are determined by first calculating the fluid and crystal chemical potentials, and then identifying the coexistence points as the intersection of these curves at fixed temperature and pressure.
For both DBI and DBN, experimental solubilities correspond to protein volume fractions of $\phi=10^{-3}$ or lower~\cite{James:2015}.
At such low-densities simple local Monte Carlo (or molecular dynamics) sampling of the fluid phase is computationally inefficient, because transport is relatively slow. While this problem can be alleviated with advanced sampling methods such as aggregation volume bias Monte Carlo~\cite{Chen:2000} and event chain Monte Carlo~\cite{Bernard:2009}, we here instead estimate the fluid properties from the second virial coefficient, $B_2$, which is calculated as in Ref.~\citen{Kern:2003} (see SI~\cite{Fusco:2013}). Because the patch energies are high, $B_2$ can become very large and negative at low temperatures, but the protein density remains sufficiently low for $|B_2\rho|\ll 1$ in the regime of interest. In order to confirm that the third virial coefficient, $B_3$, can then be neglected, we bound its value as follows. Because the patch properties are such that triply-bonded triplets cannot form, the dominant contribution to $B_3$ comes from doubly-bonded triplets. This term scales as $B_2^2$, hence $|B_3|\lesssim|B_2|^2\ll 1$ in the regime of interest, and its contribution to $\beta\mu_f$ is negligible.
Higher-order corrections are similarly expected to be negligible, thus justifying this theoretical expediency.

The fluid equation of state and chemical potential, $\mu_f$, can then be written as

\begin{gather}
	\frac{\beta p}{\rho} = 1 + B_2\rho,
\end{gather}

\begin{gather}
	\beta\mu_f = \beta\mu^\mathrm{id}+2B_2\rho = \log\Lambda^3\rho + 2B_2\rho,
\end{gather}
where $\beta\mu^\mathrm{id}=\log\Lambda^3\rho$ is the chemical potential of the ideal gas, and the thermal de Broglie wavelength $\Lambda$ is set to unity, without loss of generality.
With this formulation, we have

\begin{gather}
	\rho = \frac{-1+\sqrt{1+4B_2\beta p}}{2B_2}.
\end{gather}
If $B_2$ is positive, $\partial\beta\mu_f/\partial\rho$ is also positive. If $B_2$ is negative, $\partial\beta\mu_f/\partial\rho>0$ for $\rho<-1/2B_2$, which is always true. Thus, $\beta\mu_f$ decreases with decreasing pressure.

The crystal free energy at a given pressure and temperature is calculated using numerical simulations (see SI for simulation details). and the Frenkel-Ladd method~\cite{Frenkel:1984}, which involves thermodynamically integrating from an ideal Einstein crystal. From this reference free energy, thermodynamic integration along an isobar provides the free energy at different temperatures,

\begin{gather}
	\beta\mu_c (\beta,p) = \beta_0 \mu_c (\beta_0,p) + \int\limits_{\beta_0}^{\beta} \frac{\langle H(\beta^\prime)\rangle}{N}d\beta^\prime + \int\limits_{\beta_0}^{\beta} \beta^\prime\frac{\langle d U / d\beta^\prime \rangle}{N} d\beta^\prime,
	\label{eq:isobarIntegration}
\end{gather}
where $\langle H\rangle=p\langle V\rangle + \langle U\rangle$ is the enthalpy and $\langle \cdot \rangle$ denotes thermal averaging. Because of the highly constrained geometry of the patchy models, both crystals are almost incompressible. As a result, $\langle V \rangle$ is essentially independent of temperature. To high accuracy, we can thus write

\begin{gather}
	\int\limits_{\beta_0}^{\beta} \frac{\langle H(\beta^\prime)\rangle}{N} d\beta^\prime \approx \frac{1}{N}\int\limits_{\beta_0}^{\beta} \frac{\langle U(\beta^\prime)\rangle}{N} d\beta^\prime + \frac{p}{\rho} (\beta-\beta_0).
	\label{eq:isobarIntegrationApprox}
\end{gather}
Note that at sufficiently low pressures, the second term is also negligible.

\begin{figure}
	\includegraphics[width=\textwidth]{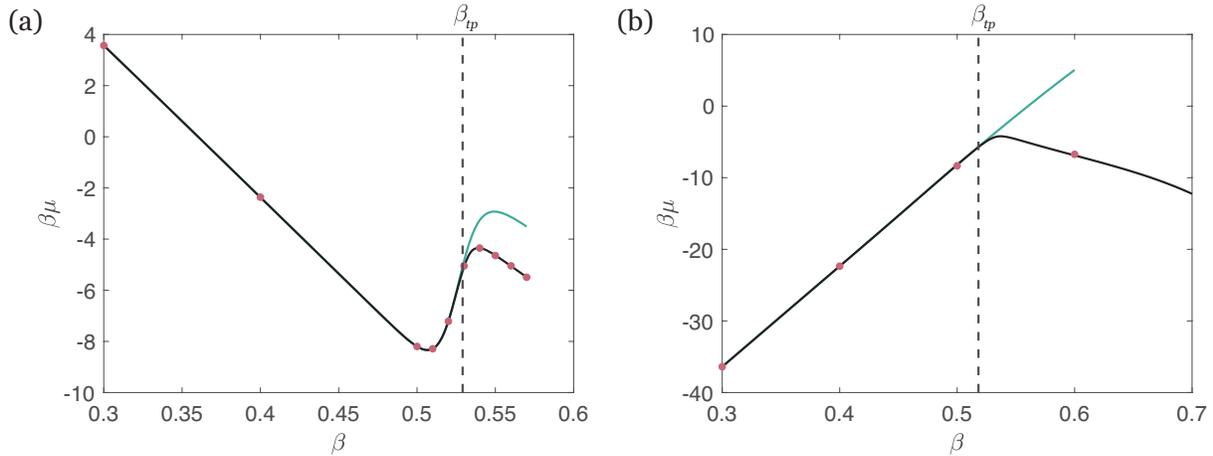}
	\caption{Calculated $\beta\mu_c$ using thermodynamic integration starting from $\beta=0.3$. The simulation data (black) fully matches the individual Einstein crystal simulations (red data points). Estimates of $\beta\mu_c$ (blue) become significantly flawed at low temperatures, but because this regime is beyond the triple point, $\beta_\mathrm{tp}$ (dashed line), the DBI solubility line is unaffected. (a) Patch 4 is deactivated below $T_a$, with $\tau=0.05$, and (b) Patch 4 energy follows the Wentzel-Gunton model with $\Delta s_w=-50$.}
	\label{fig:betamuc}
\end{figure}

We further approximate that all the crystal bonds are active, and hence $\langle U(\beta)\rangle \approx U_0(\beta)$, where $U_0(\beta)$ is the ground state energy, and $\langle dU/d\beta \rangle\approx dU/d\beta$. 
While this approximation is generally quite good, it is overly crude in the patch deactivation regime, where the patch energy decreases rapidly around $\beta_a$, and vanishes when temperature is reduced further. As a result, $\langle dU/d\beta\rangle\ll dU/d\beta$, which can result in a significant correction to $\beta\mu_c$ (see Fig.~\ref{fig:betamuc}a). In the Wentzel-Gunton model, the patch similarly becomes non-attractive for $\beta>\beta_\mathrm{ref}$, and upon further lowering the temperature, it eventually becomes repulsive. The topology of the DBI crystal then changes 
and the energy of the crystal once again becomes temperature-independent, which leads to a bending of the evolution of the chemical potential with temperature (Fig.~\ref{fig:betamuc}b). In both cases, however, the DBI solubility curve is unaffected, because these changes occur in a region where DBI is metastable with respect to DBN.

Under this approximation, the chemical potential of the crystal for the MLG model can be written as

\begin{gather}
	\beta\mu_c(\beta,p) \approx \beta_0\mu_c(\beta_0,p) + \frac{p}{\rho} (\beta-\beta_0) + \xi(\beta),
	\label{eq:betamuS2005}
\end{gather}
where

\begin{gather}
	\xi(\beta)
	=\int\limits_{\beta_0}^\beta \Big[-\varepsilon_\mathrm{tot}+\Delta\varepsilon(1)n_w - 2n_w \varepsilon_w  + 2n_w(\frac{d\Delta s_w}{d\beta^\prime} - \beta^\prime\frac{d\varepsilon_w}{d\beta^\prime})\Big] d\beta^\prime\\
	= (\Delta\varepsilon(1) n_w - \varepsilon_\mathrm{tot})(\beta-\beta_0) + 2n_w\Big[ (\Delta s_w(\beta) - \Delta s_w(\beta_0))- (\beta \varepsilon_w(\beta)-\beta_0 \varepsilon_w(\beta_0)) \Big].
\end{gather}
We thus have that

\begin{gather}
	\frac{\beta\mu_c}{d\beta} \approx (\Delta\varepsilon(1)n_w-\varepsilon_\mathrm{tot}) + 2n_w \Big[\frac{d\Delta s_w}{d\beta}-\varepsilon_w-\beta\frac{d\varepsilon_w}{d\beta}\Big],
\end{gather}
which has a minimum when

\begin{gather}
	\Gamma(\beta)\equiv \frac{d\Delta s_w}{d\beta}-\varepsilon_w-\beta\frac{d\varepsilon_w}{d\beta} = \frac{\varepsilon_\mathrm{tot}}{2n_w}-\frac{\Delta\varepsilon(1)}{2}.
	\label{eq:gS2005}
\end{gather}
As already noted, $\beta\mu_f$ decreases with decreasing pressure, and because by thermodynamic stability so does $\rho$, an inverted solubility regime is only obtained when the slope of $\beta\mu_c$ with respect to $\beta$ is positive.
For $\Gamma(\beta)>\varepsilon_\mathrm{tot}/(2n_w)-\Delta\varepsilon(1)/2$, the slope of $\beta\mu_c$ is positive, hence inverted solubility is observed.

For the Wentzel-Gunton model, the change in $\beta\mu_c$ with temperature can be similarly estimated. We can write the energy per particle in the crystal as

\begin{gather}
	U(\beta)/N = -\varepsilon_\mathrm{tot} - 2\Delta s_w \Big(\frac{1}{\beta_\mathrm{ref}}-\frac{1}{\beta}\Big).
\end{gather}
and hence, following Eq.~\eqref{eq:isobarIntegration},

\begin{gather}
	\beta\mu_c(\beta,p)
	= \beta\mu_c(\beta_0,p)+\Big(-\varepsilon_\mathrm{tot}-\frac{2\Delta s_w}{\beta_\mathrm{ref}}-\frac{p}{\rho}\Big) \Big(\beta-\beta_0\Big).
	\label{eq:betamucSimple}
\end{gather}
The slope of $\beta\mu_c$ with respect to $\beta$ is positive when $-\varepsilon_\mathrm{tot}-p/\rho>2\Delta s_w/\beta_\mathrm{ref}$, thus resulting in inverted solubility.

Although  it is not possible to write a compact expression for $\beta\mu_c$ for the temperature-(de)activated patchy model--the associated integrals need to be evaluated numerically-- the phenomenology is similar. The solubility is inverted in the region where $\beta\mu_c$ has a positive slope, i.e., around $T_a$. The associated jump in $\beta\mu_c$ due to patch deactivation with increasing $\beta$ can be seen in Fig.~\ref{fig:betamuc}a.

If patch energies are modified by either randomly perturbing them or by scaling them by a constant factor, the free energy of this altered model is estimated from the original model, assuming that the crystal free energy can be expressed as

\begin{gather}
	\beta A^\prime = \beta A - \beta U_0/N + \beta U_0^\prime/N,
\end{gather}
where $A^\prime$ is the Helmholtz free energy and $U_0^\prime$ is the ground state crystal energy for altered model. This treatment amounts to neglecting the change in crystal entropy upon weakening or strengthening the patches, which is a small contribution. We also verify that the crystal remains stable at the temperatures of interest.

The approximations described above allow the expedited consideration of coexistence points that constitute the solubility curves by generating $\beta\mu_f$ and $\beta\mu_c$ as functions of $\beta$ at various pressures. 

\section{Inverted Solubility from Hydrophobicity Models}
\label{subsub:nw}
In order for the microscopic hydrophobicity models described above to give rise to solubility inversion, a sufficient number of water molecules need to be involved. In this section we first consider physical bounds on that number, and then consider how the corresponding crystal solubility lines are affected.

\subsection{Effect of Parameters on Solubility Lines}
The key free parameter in hydrophobicity models is the number of water molecules solvating the hydrophobic patch, $n_w$. We first estimate the number of water molecules possibly available around Patch 4, by calculating the SASA for the participating residues~\cite{Khan:2019} and then compute

\begin{gather}
	n_w = A_{4} \rho_w\int\limits_{3\angstrom}^{4.5\angstrom} g_C(r)dr,
\end{gather}
where $A_{4}$ is the solvent accessible surface area of Patch 4, $\rho_w=3.3\times10^{-2}\angstrom^{-3}$ is the number density of water in the bulk at room temperature, and $g_C(r)$ is the radial distribution function of water around carbon atoms determined in Ref.~\citen{Altan:2018}. This estimate thus assumes that (i) the solvent has a radius of $1.4\angstrom$ (the SASA definition), (ii) the average van der Waals radii of protein heavy atoms is $\sim 1.6\angstrom$, and (iii) the first solvation shell ends with the first peak of $g(r)$ at $4.5\angstrom$. We further assume that the measured surface is flat, which is here but a small correction.
If we furthermore assume that all residues contributing to Patch 4 are hydrophobic, then $n_w=133-140$ for all six protein structures. However, because Patch 4 contains only a handful of hydrophobic residues a more realistic estimate should decrease $A_{4}$. Assuming that a residue is hydrophobic if it is labeled as such in \emph{any} of the hydrophobicity scales considered in Sec.~\ref{sec:surfHydro} gives instead $n_w=43-48$. 
Because the hydrophobic residues within Patch 4 do not constitute a contiguous area, the configuration of contributing water molecules solvating them will additionally be affected by the nearby hydrophilic surface residues. This estimate should thus be treated as an upper bound. 
Note that the P23T mutation does not seem to be associated with a systematic change in $A_4$, and thus $n_w$.

We compare this result with the number of water molecules needed for Patch 4 to have its measured bond strength. In particular, if we attribute the entire Patch 4 energy to the change in free energy upon moving solvating water molecules to the bulk, then the MLG model gives $\varepsilon_4=n_w\Delta \varepsilon(1)$, and thus $n_w\sim23$. Because multiple hydrogen bonds also contribute, however, this number should also be treated as an upper bound, that is consistent with yet tighter than the above bound. 

\begin{figure}
	\includegraphics[width=0.5\textwidth]{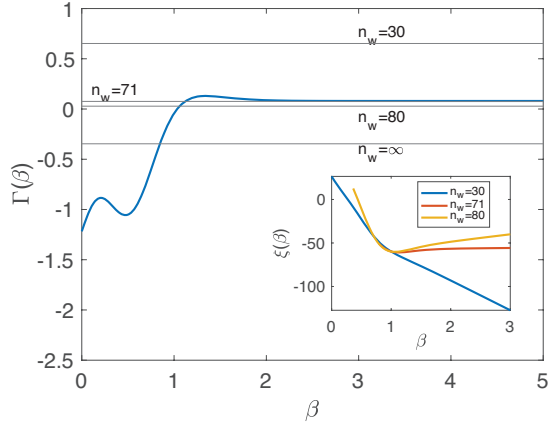}
	\caption{The minimum of $\xi(\beta)$, and hence of $\beta\mu_c$, is obtained by the intersection of $n_w$ values (black lines) with $\Gamma(\beta)$ (the blue curve) as given in Eq.~\eqref{eq:gS2005}. The inset shows the corresponding $\xi(\beta)$, i.e. the temperature-dependent part of $\beta\mu_c$ for various $n_w$ using model parameters reported by Silverstein et al. Here, 30 water molecules are not sufficient to invert solubility, but $n_w^*\geq 71$ is. }
	\label{fig:fplotS2005}
\end{figure}

We can now contrast these bounds with the minimum number of water molecules, $n_w^*$, that need to be displaced to invert solubility. For the MLG model,  we use Eq.~\eqref{eq:gS2005} and the sum of DBI patch energies, $\varepsilon_\mathrm{tot}=60$, to  estimate 
$n_w^*$; it must be such that $\xi(\beta_\mathrm{min})$ is a minimum, i.e., $\Gamma(\beta)>\varepsilon_\mathrm{tot}/(2n_w)-\Delta\varepsilon(1)/n_w$. In other words, the solubility is inverted if $\beta>\beta_\mathrm{min}$. The numerical solution in Fig.~\ref{fig:fplotS2005} shows that $n_w^*\gtrsim 71$. The corresponding change in $\xi(\beta)$ is given in the inset. It should be noted, however, that $n_w^*$ depends on the MLG model parameters. For the multiplicities of Shiryayev et al., for instance, inverted solubility is possible with a mere $n_w^*\sim18$. 
These degeneracies, however, are fairly arbitrary~\cite{Moelbert:2003}.  
Choosing $q_\mathrm{ob}=10$ and $q_\mathrm{os}=1$, in particular, seems unphysical, because the ordered bulk degeneracy is unlikely to be an order of magnitude larger than that of the ordered shell. We thus expect $n_w^*\gtrsim 71$ to be physically more reasonable.

\begin{figure}
	\includegraphics[width=0.5\textwidth]{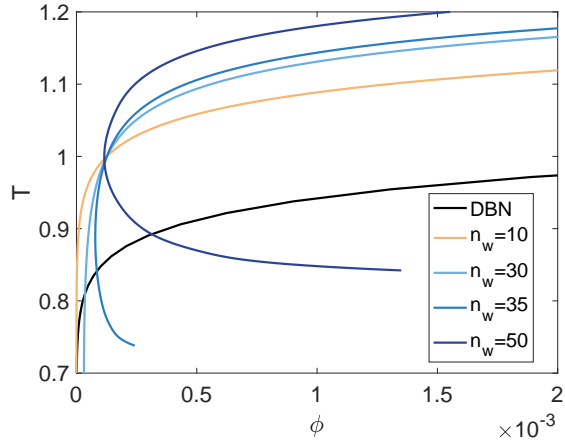}
	\caption{Solubility lines corresponding to different values of $n_w$ for $\varepsilon_\mathrm{tot}^\prime = 30$, for which $n_w^*\geq 35$.}
	\label{fig:s2005nw}
\end{figure}

However, because $\varepsilon_\mathrm{tot}=60$ results in room temperature solubilities that are orders of magnitude lower than their experimental counterpart, and in light of the various sources error in patch energy determination (Sec.~\ref{subsubsec:patchyModel}), Ref.~\citen{Khan:2019} proposed a reasonable correction would be to halve patch energies. For $\varepsilon_\mathrm{tot}=30$, we have $n_w^*\sim35$ (Fig.~\ref{fig:s2005nw}), which is less than the 40 or so water molecules solvating hydrophobic residues in Patch 4, but more than the energetic estimate. 

In light of the many estimates involved in the above analysis, the hydrophobic effect as a cause of inverted solubility, although weakly supported, cannot be eliminated outright. Even if the hydrophobicity model parameters are kept constant, a possible resolution could be for Patch 4 to be stronger than estimated and the other patches weaker. The hydrophobicity scenario, however, does severely constrain the patch model parameters.

\subsection{Solubility Lines for Models of Hydrophobicity}
\label{subsub:WG2008S20005}
In the previous subsection we determined that the hydrophobicity scenario for inverting solubility requires a fine balance between the protein-protein patch energies, the size of the hydrophobic patch, and the number of water molecules solvating it. While this rare confluence of factors could explain why inverted solubility is not common among proteins, it is natural to wonder whether the presence of weak hydrophobic patches, which are ubiquitous in proteins, affects solubility lines without engendering a regime of inverted solubility. In this section, we study the Wentzel-Gunton model in order to examine this possibility.

\begin{figure}
	\includegraphics[width=0.5\textwidth]{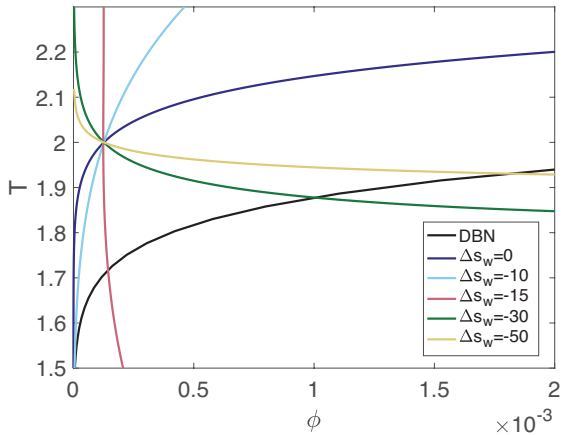}
	\caption{ Larger magnitudes of $\Delta s_w$ invert solubility, whereas $\Delta s_w=-15$ results in solubility that only weakly depends on temperature, and $\Delta s_w=-10$ (light blue) results in normal solubility. Note that, for the latter case, even though the solubility is not inverted, the solubility line is markedly altered compared to the $\Delta s_w=0$ case (dark blue).}
	\label{fig:WG2008_fig}
\end{figure}

The solubility lines for the Wentzel-Gunton model in Fig.~\ref{fig:WG2008_fig} are specifically obtained for $\beta_\mathrm{ref}=0.5$, to match the experimental solubility as in Ref.~\citen{Khan:2019}, but our observations are qualitatively independent  of this choice. 
Setting $\Delta s_w=-10$, which is here akin to $n_w\sim 20$ (assuming that the temperature-dependent energy in the MLG model scales as $n_w$), results in normal solubility, but the steepness of the solubility curve changes markedly compared to $\Delta s_w=0$. Setting $\Delta s_w = -15$ ($n_w\sim30$) results in the DBI solubility being almost independent of temperature and in DBN being more stable than DBI at $T<T_\mathrm{tp}\sim 1.7$. Further reducing $\Delta s_w$ 
results in an inverted solubility regime. The solubility curve then flattens below $T\sim 2$ and $\phi_\mathrm{tp}$ moves to higher packing fractions.  
These observations thus reveal that the presence of an inverted solubility regime is the limit case of a continuum of how hydrophobicity impacts the solubility line.

\section{Solubility Lines for Temperature Deactivated Patches}
\label{subsubsec:tempDeacResults}
Absent clear microscopic evidence for the hydrophobic effect, we finally consider a generic model for patch deactivation. The temperature-deactivated patchy model, which was used to successfully capture the inverted solubility of DBI~\cite{Khan:2019}, stabilizes the crystal with increasing temperature without referring to any specific microscopic mechanism.  
In this section, we first discuss the physical constraints on the model parameters and then consider how solubility lines change with model parameters, paying particular attention to the robustness of the inverted solubility regime. We also estimate the binodal and the critical temperature, which has been experimentally for some human $\gamma$D-crystallin variants~\cite{Mcmanus:2007}.

\subsection{Parameter Estimates}
Despite the absence of an explicit microscopic interpretation for the (de)activation model, one can still place some reasonably solid physical constraints its tuning parameters. 
First, the (de)activation temperature $T_a$ must lie in the vicinity of the triple point, and thus $T_a\sim T_{\mathrm{tp}}$. For our model, the choice $T_a=1.9$ ensures that the deactivation of Patch 4 makes  DBI  metastable with respect to DBN for $T<T_\mathrm{tp}$. 
Second, $\tau$, which sets the temperature range over which (de)activation takes place, ought to capture the degree of cooperativity of the underlying microscopic effect. It therefore cannot be arbitrarily small, as it would be at a thermodynamic phase transition. Denaturing a protein, for instance, takes place over a few degrees, and any smaller scale rearrangement that involves tens to hundreds of atoms should spread over at least $\gtrsim10K$. However, if this change takes place too gradually, say over $\gg10K$, then inverted solubility cannot be observed for typical protein-protein interactions.
We thus here consider a temperature range of $\sim 10K$, which corresponds to setting $\tau=0.05$. 

\begin{figure}
	\includegraphics[width=\textwidth]{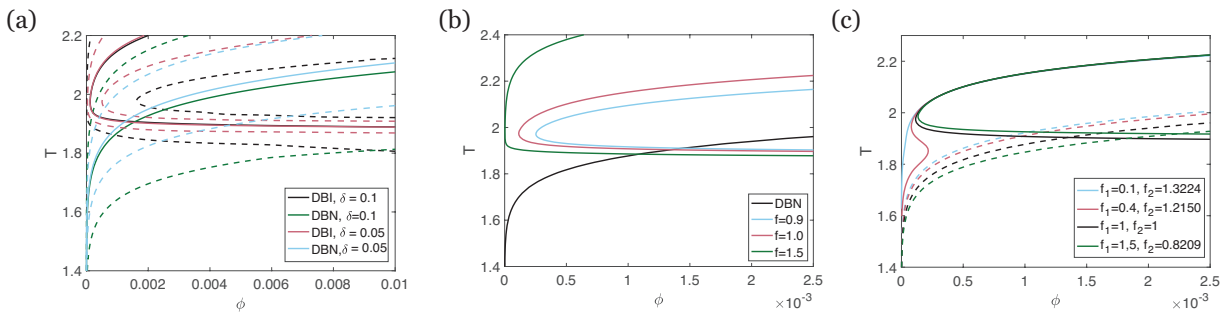}
	\caption{(a) Average solubility lines for perturbed parameters. Dashed lines denote $95\%$ confidence intervals. DBI solubility line for $10\%$ error (black) and $5\%$ error (red), as well as DBN solubility line for $10\%$ error (green) and $5\%$ error (blue) are shown. Higher error levels increase the uncertainty in $\phi_\mathrm{tp}$, as well as the minimum solubility observed for DBI, but inverted solubility is maintained. (b) The effect of changing the energy of the temperature-deactivated patch, such that $\varepsilon_4^\prime=f\varepsilon_4$. (c) The effect of changing $\varepsilon_4$ but keeping the total patch energies of DBI constant.}
	\label{fig:tanh_fig1}
\end{figure}

We first investigate how varying the patch energies impacts the phase diagram, keeping $T_a=1.9$ and $\tau=0.05$ constant. As previously reported~\cite{Khan:2019}, the resulting phase diagram (Fig.~\ref{fig:tanh_fig1}a) exhibits a re-entrance regime bounded by the DBI solubility line, as well as a triple point between the fluid and the two crystal forms.
The solubility lines that result from perturbing the patch energies by $5\%$ and $10\%$ are shown in  Fig.~\ref{fig:tanh_fig1}a. Although the solubility lines then shift to substantially lower or higher densities, the existence of an inverted solubility regime is robust. The errors inherent to the overall parameterization of the model are therefore qualitatively benign.

We next investigate the robustness of the results with respect to the relative strength of the temperature-deactivated fourth patch, $\varepsilon_4$. This question is of interest for two main reasons: (i) the strength and robustness of solubility inversion depend sensitively on the strength of that patch; and (ii) the ordering of the single mutant solubilities directly correlates with their respective Patch 4 energies.

We first multiply $\varepsilon_4$ by $f\in\{0.9,1.0,1.5\}$, while keeping the other patch parameters constant (Fig.~\ref{fig:tanh_fig1}b). Increasing the strength of Patch 4 
systematically decreases the solubility of DBI as well as $\phi_\mathrm{tp}$. Interestingly, the decrease in solubility with increasing $f$ is consistent with the experimental observations for the single mutants, P23T, P23S, and P23V~\cite{Mcmanus:2007}. 
Because a stronger Patch 4 decreases the DBI solubility (R36S+P23T double mutant), assuming that the difference between crystals will arise due to Patch 4 only, we speculate that if two other double mutants, R36S+P23S and R36S+R23V, were crystallized with similar crystal contacts, then their inverted solubility would have a similar ordering. 

We next change the energy of the temperature-dependent patch while keeping the total energy of DBI patches constant, i.e., $\varepsilon_\mathrm{tot}=f_1\varepsilon_4 + f_2(\varepsilon_1+\varepsilon_2+\varepsilon_3+\varepsilon_5)$ (Fig.~\ref{fig:tanh_fig1}c). (Because the second patch corresponds to a shared contact between DBI and DBN, the DBN solubility is then also slightly perturbed.) As in the first case, the inverted solubility regime vanishing upon markedly reducing the strength of Patch 4.  The difference is that DBN is here metastable with respect to DBI within the probed temperature range, while DBI becomes metastable with respect to DBN otherwise. 
For $f_1=0.4$, DBN is still metastable with respect to DBI, but the narrow range of inverted solubility is then replaced by standard solubility at both lower and higher temperatures.

\begin{figure}
	\includegraphics[width=\textwidth]{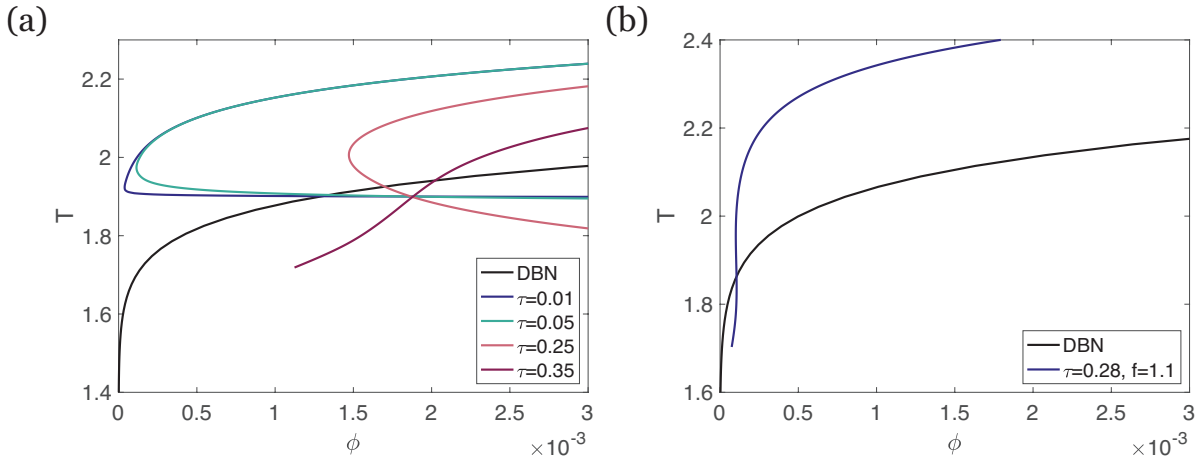}
	\caption{(a) As $\tau$ is increased, the DBI solubility line becomes less flat, and eventually inverted solubility is lost ($\tau=0.35$). (b) Manipulating the sum of DBI patch energies and $\tau$, one can show that it is possible to have a temperature range within which the solubility is almost temperature-independent.}
	\label{fig:tanh_fig2}
\end{figure}

We finally investigate the robustness of the phenomenology with respect to changes in $\tau$.
Decreasing $\tau$ corresponds to a faster temperature (de)activation of the patch, which flattens the inverted solubility region and results in $T_\mathrm{tp}\rightarrow T_a$ as $\tau \rightarrow 0$ (Fig.~\ref{fig:tanh_fig2}a). $\phi_\mathrm{tp}$ similarly gets pushed to higher packing fractions, suggesting that a protein solution prepared very near $T_a$ could reach remarkably high concentrations compared to solutions prepared at surrounding temperatures. However, as argued above, very small values of $\tau$ are not here physically meaningful.
Conversely, increasing $\tau$ weakens this transition and eventually eliminates the solubility inversion regime.
Increasing $\tau$ also renders the inverted solubility less robust with respect to varying patch energies. 
Interestingly, a specific choice of $\tau$, with a minor tweak to patch energies ($\tau=0.28$, $\varepsilon_i^\prime=1.1\varepsilon_i$), gives rise to a nearly vertical solubility curve (Fig.~\ref{fig:tanh_fig2}b), similar to the temperature-independent solubility of apoferritin~\cite{Vekilov:2002}.

\subsection{Estimation of the Critical Temperature}
\label{sec:PD}
Although various theoretical results suggest that a closed-loop binodal with multiple critical points is possible upon introducing temperature-dependent binding energies~\cite{Shiryayev:2005,Wentzel:2008,de:2016}, no experimental evidence of such a closed-loop binodal is found for any human $\gamma$D-crystallin mutant. In addition, experiments find that the P23V mutation, which also inverts solubility, has a binodal indistinguishable of that of the wild type~\cite{Mcmanus:2007}.
We thus estimate the liquid-liquid binodal and the associated critical temperature, $T_c$, to determine if the deactivation of Patch 4 affects the solution properties.  Here we use Wertheim's perturbation theory~\cite{Wertheim:1984,Wertheim:1984b}, which provides quantitatively good estimates of the binodals in patchy models (see SI for details~\cite{Liu:2009}).

\begin{figure}
	\includegraphics[width=0.5\textwidth]{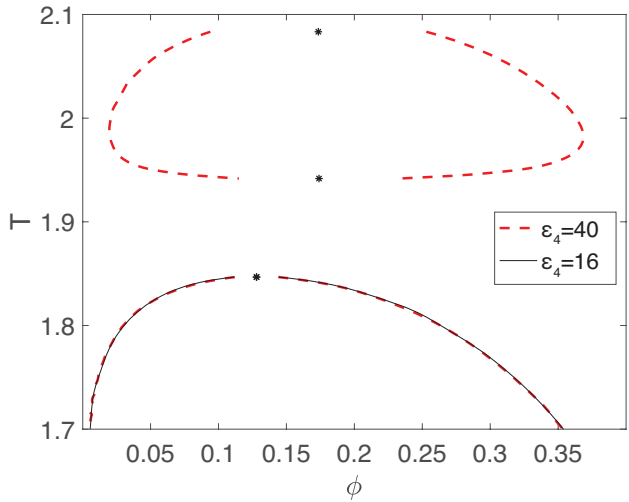}
	\caption{The liquid-liquid binodal regions for the temperature-deactivated patch model for various Patch 4 energies.  For the reference model patch energy, $\varepsilon_4=16$ ($\tau=0.05$, $T_a=1.9$), there is no closed-loop binodal. Only if the energy of Patch 4 is greater than $\varepsilon_4>36$, here for illustration $\varepsilon_4=40$ (red), are multiple critical points and a closed-loop binodal obtained.}
	\label{fig:binodals}
\end{figure}

Choosing $\tau=0.05$ and $T_a=1.9$, as above, results in a typical binodal with a single critical point at $T_c=1.85$ (Fig.~\ref{fig:binodals}). Hence, without altering patch energies, we do not observe any signature of a closed-loop binodal in our model (Fig.~\ref{fig:binodals}), which is consistent with experimental results~\cite{Mcmanus:2007}.
In order to determine how far our model is from exhibiting a closed-loop binodal, we systematically increase the Patch 4 energy. Only above $\varepsilon_4>36$, does a closed-loop binodal appear. See example in Fig.~\ref{fig:binodals}.
If only the energy of Patch 4 is more than doubled and is disproportionately stronger than any other patch in the model does this phenomenon appear. This perturbation falls far outside of the error estimates for patch energies, which supports the robustness of our model prediction.

\section{Conclusion}
\label{sec:conclusions}
We have here attempted to rationalize the inverted solubility of certain mutants of $\gamma$D-crystallin based on microscopic models of protein-protein interactions and their temperature dependence. We have paid particular attention to the putative role of hydrophobicity. Estimating surface hydrophobicity using different scales did not reveal the presence of any pertinent surface feature, but microscopic models of hydrophobicity suggest that the amount of available surrounding water molecules might be close be sufficient. Although our analysis falls short of concluding whether hydrophobicity plays a determining role,  the scenario nonetheless seems a bit far fetched. 
A more convincing determination would likely require for the water structure around the region of interest to be more specifically probed. 
Because numerical simulations of standard model of water are insufficiently sensitive to this feature~\cite{Altan:2018} to be conclusive, simulations with more sophisticated water models~\cite{Kessler:2015,Tainter:2015}, and neutron diffraction or hydrogen-deuterium exchange experiments might be more productive avenues. 

Even though the microscopic origin of inverted solubility in human $\gamma$D-crystallin thus still remains somewhat elusive, additional insight from the crystallization of other double mutants, such as R36S+P23V and R36S+P23S, might be helpful in identifying generic features that might have eluded the analysis thus far. 
Repeating the above structural and thermodynamical for these mutants could help tease out more subtle features.

\begin{acknowledgement}
\label{sec:acknowledgements}
We wish to thank our colleague David Beratan for his continued support. It has been a pleasure to exchange fruits and science with him over the last decade. We acknowledge support from National Science Foundation Grant no.~NSF DMR-1749374. This work used the Extreme Science and Engineering Discovery Environment (XSEDE), which is supported by National Science Foundation grant number ACI-1548562, as well as the Duke Compute Cluster. 
We warmly thank Diana Fusco for stimulating discussions as well as Amir Khan for discussing and sharing experimental data ahead of publication.
Data relevant to this work have been archived and can be accessed at \url{https://doi.org/10.7924/XXXXXXX}. 
\end{acknowledgement}

\begin{suppinfo}
The Supporting Information is available free of charge on the \href{http://pubs.acs.org}{ACS Publications website} at DOI: \url{10.1021/XXXX}.
\end{suppinfo}

\section*{Supporting Information}

\subsection*{Analyzing Crystal Structures}

The residues identified to be involved in Patch 4 are: 

\vspace{5mm}

\noindent LYS 2, \textbf{TYR 16}, GLU 17, SER 19, SER 20, ASP 21, HIS 22, PRO/THR 23, ASN 24, \textbf{PRO 27}, \textbf{TYR 28}, \textbf{PRO 48}, ASN 49, \textbf{TYR 50}, \textbf{PRO 82}, HIS 83, SER 84, SER 86, ARG 88, \textbf{TYR 97}, \textbf{VAL 125}, \textbf{LEU 126}, GLU 127, GLY 128, SER 129, \textbf{MET 146}, \textbf{PRO 147}, GLY 148, ASP 149, \textbf{TYR 150}, ARG 151, ARG 152, GLN 154, ASP 155.

\vspace{5mm}

\noindent Residues in bold are deemed hydrophobic in at least one hydrophobicity scale. We compare the solvent accessible areas (SASA) obtained from these selections, $A_4$ , across the available mutant crystal structures to evaluate if a trend in hydrophobicity that could explain the difference between normal and inverted solubility could be measured. If the P23T mutation were to increase the SASA, then maybe a tipping $n_w$ could be reached. The bare value of $A_4$ gives R36S $>$ DBI$^*$ $>$ R58H $>$ WT $>$ DBN$^*$ $>$ P23T$^*$, where the asterisk denotes structures that contain the solubility inverting mutation. If we consider only the hydrophobic residue contribution to Patch 4, we instead have R58H $>$ R36S $>$ DBI$^*$ $>$ DBN$^*$ $>$ WT $>$ P23T$^*$. In any case, no systematic change is detected, thus further weakening the case for hydrophobicity as key to solubility inversion.

\subsection*{Model Parameters and Geometry}
Model parameters are obtained from all-atom molecular dynamics (MD) simulations using Gromacs~\cite{Berendsen:1995} versions 5.1 and 2016. 
Patch energies are chosen as the depth of the well in the PMF, after zeroing its long distance energy value. The interaction range for each patch was then calculated by matching the second virial coefficient, $B_{2,\alpha\beta}$, of the square well potential to that of the PMF

\begin{gather}
	B_{2,\alpha\beta} = -\frac{1}{2} 4 \pi \int \Big(e^{-\beta U(r)}-1\Big)r^2dr,
\end{gather}
from which the interaction range is found to be

\begin{gather}
	\lambda_{\alpha\beta} = \Bigg(\frac{3\int(e^{-\beta U_\mathrm{PMF}(r)}-1)r^2dr}{e^{\beta\varepsilon_{\alpha\beta}}-1}+1\Bigg).
\end{gather}

The details of the all-atom simulations resulting in model parameters are given in the supporting information of Ref.~\citen{Khan:2019}. The resulting model parameters are summarized in Table~\ref{tbl:params}. Contacts are labeled cY, where Y is a Roman numeral for DBN contacts, and an Arabic numeral for DBI contacts. Contact 2 (c2) of DBI and contact I (cI) of DBN are very similar in terms of geometry and the nature of their interactions and were hence merged. Contact V (cV) and VII (cVII) of DBN are also almost identical and were also merged. 

The patch locations and dihedral angles are listed in Table~\ref{tbl:modelGeom}. A number of modifications were made to the geometry of the actual protein crystal to be able to represent the proteins as spherical objects in an orthorhombic unit cell and are detailed in the SI of Ref.~\citen{Khan:2019}. Here we briefly quote the resulting geometry (Table~\ref{tbl:modelGeom}), and note that we do not expect these topological changes to qualitatively affect our results, as it was previously shown that patch locations have only a negligible effect on the phase diagram topology~\cite{Fusco:2013}. Note that the resulting patch geometry is such that dihedral constraints are necessary to properly separate the crystal ground states within the energy landscape. 

\begin{table}
\caption{Patch parameters obtained from all-atom molecular dynamics simulations and slightly modified to adjust the crystal geometry}	
	\centering
\begin{tabular} {c || c | c | c | c | c}
	Contact & $\cos\delta_\alpha$ & $\cos\delta_\beta$ & $\Delta\phi_{\alpha\beta}$ (rad) &$\varepsilon$ ($k_\mathrm{B}T$) & $\lambda$ ($\sigma$) \\\hline\hline
	c1 & 0.956 & 0.963 & 0.07 & 21.5 & 1.025 \\ \hline
	c2 & 0.935 & 0.935 & 0.21 & 6.3 & 1.1 \\ \hline
	c3 & 0.992 & 0.994 & 0.14 & 10.3 & 1.06 \\ \hline
	c4 & 0.997 & 0.997 & 0.06 & 15.8 & 1.025 \\ \hline
	c5 & 0.947 & 0.966 & 0.13 & 6.0 & 1.037 \\ \hline
	cI &0.972	&0.981&	0.21&	6.3	&1.1 \\ \hline
	cII &	0.96 & 0.96 & 0.3 &	8.7	& 1.11 \\ \hline
	cIII &	0.95	& 0.95 &	0.2&	10.1&	1.1 \\ \hline
	cIV	& 0.96	&0.937&	0.3&	4.3&	1.14 \\ \hline
cV	& 0.95	& 0.95	& 0.21	& 18.5	& 1.153 \\ \hline
cVI	&0.898	&0.927	& 0.3	&8.6	&  1.11 \\ \hline
cVII	& 0.95 & 0.95 & 0.21 &	18.5 &	1.53 \\ \hline
\end{tabular}
\label{tbl:params}
\end{table}

\begin{table}
\caption{Vectors defining each patch location and dihedral angle, $\phi_{\alpha\beta}$. Note that because of geometric constraints of the DBN structure, contacts cII, cIV, and cVI are collocated.}
	\centering
\begin{tabular} {c || c | c| c| c}
	Contact & $\hat{e}_1$ & $\hat{e}_2$ & $\hat{e}_3$ & $\phi_{\alpha\beta}$\\ \hline\hline
	c1 &	0.6614 	&0.6543 &0.3667 &-2.2703 \\
	   & 0.6473 & -0.2853 & 0.7068 & \\ \hline
	c2/cI & -0.9854 & 0.0471 & 0.1101 & 0.0 \\ 
	   & 0.9854 & -0.0471 & -0.1101 &  \\ \hline
	c3 & -0.0141 & -0.9402 & 0.3403 & 0.0 \\
	   & 0.0141 & 0.9402 & -0.3403 & \\ \hline
	c4 & 0.3231 & 0.1911 & -0.9269 & -2.1288 \\ 
	   & 0.3089 & -0.7486 & -0.5867 & \\ \hline
	c5 & -0.6474 & 0.2851 & -0.7068 & 2.2698 \\ 
	   & -0.6613 & -0.6543 & -0.3669 & \\ \hline
	cII* & -0.9494 & 0.3101 & -0.0499 & -2.4946 \\
		& 0.0011  & 0.4334  & 0.9012 & \\ \hline
	cIII & -0.6455 & -0.5123 & -0.5665 & -1.80 \\
	    & -0.7102 & -0.5524 & 0.4365 & \\ \hline
	cIV* & -0.9494 & 0.3101 & -0.0499 & -2.4946 \\
		& 0.0011  & 0.4334  & 0.9012 & \\ \hline
	cV/VII & 0.3162 & -0.8026 & -0.5058 & -2.1529 \\ 
	     & 0.4927 & 0.8531 & 0.1716 & \\ \hline
	cVI* & -0.9494 & 0.3101 & -0.0499 & -2.4946 \\
		& 0.0011  & 0.4334  & 0.9012 & \\ \hline
\end{tabular}
\label{tbl:modelGeom}
\end{table}

\subsection*{Determining Solubility Lines}

For the patchy model studied here, the second virial coefficient, $B_2$, is given as

\begin{gather}
	B_2 = -\frac{1}{2V}\int d\mathbf{r}\int d\chi \int d\mathbf{\Omega}_1 \int d\mathbf{\Omega}_2  (e^{-\beta u(|\mathbf{r}|,\mathbf{\Omega}_1,\mathbf{\Omega}_2)}-1),
\end{gather}
where the integrand is the Mayer function, and the integration is performed over particle positions, $\mathbf{r}$, orientations $\mathbf{\Omega}$, and dihedral $\chi$. Computing this integral for our patchy model gives

\begin{gather}
	\frac{B_2}{B_\mathrm{HS}} = 1 - \sum (\lambda_{ij}^3-1)\sin^2\Big(\frac{\delta_i}{2}\Big)\sin^2\Big(\frac{\delta_j}{2}\Big)(e^{\beta\varepsilon_{ij}}-1)(2\Delta\phi_{ij}) \\
	- (\lambda^3_{IV}-\lambda^3_{II})(e^{\beta \varepsilon_{IV}}-1)\sin^2\Big(\frac{\delta_{IV,\alpha}}{2}\Big)\sin^2\Big(\frac{\delta_{IV,\beta}}{2}\Big)(2\Delta\chi)\\
	-(\lambda_{II}^3-1)(e^{\beta(\varepsilon_{II}+\varepsilon_{IV}+\varepsilon_{VI})}-1)\sin^2\Big(\frac{\delta_{II,\alpha}}{2}\Big)\sin^2\Big(\frac{\delta_{II,\beta}}{2}\Big)(2\Delta\chi)\\
	-(\lambda_{VI}^3-1)(e^{\beta(\varepsilon_{IV}+\varepsilon_{VI})})\sin^2\Big(\frac{\delta_{IV,\alpha}}{2}\Big)\sin\Big(\frac{\delta_{IV,\beta}+\delta_{IV,\alpha}}{2}\Big)\sin\Big(\frac{\delta_{IV,\beta}-\delta_{IV,\alpha}}{2}\Big)(2\Delta\chi)\\
	-(\lambda_{VI}^3-1)(e^{\beta\varepsilon_{VI}}-1)(2\Delta\chi)\Big[ \sin^2\Big(\frac{\delta_{IV,\alpha}}{2}\Big)\sin\Big(\frac{\delta_{IV,\alpha}+\delta_{IV,\beta}}{2}\Big)\sin\Big(\frac{\delta_{IV,\alpha}-\delta_{IV,\beta}}{2}\Big)\\
	+\sin\Big(\frac{\delta_{VI,\alpha}-\delta_{IV,\alpha}}{2}\Big)\sin\Big(\frac{\delta_{VI,\alpha}+\delta_{IV,\alpha}}{2}\Big)\sin^2\Big(\frac{\delta_{VI,\beta}}{2}\Big)\Big],
\end{gather}
where the sum is over all patch pairs excluding c$_{II}$, c$_{IV}$, and c$_{VI}$, which are collocated. The terms after the sum specifically describe the contribution of these patches. Note that $2\Delta\chi=0.6$ is the width of range of dihedral angles for these three patches, and $\Delta\phi_{V}=1.0632$ because there are four valid dihedral angles for cV with overlapping ranges.

Specialized MC simulations and estimates using the second virial coefficient were used to determine solubility lines. MC simulations have 200,000 to 1,000,000 MC sweeps, each of which consists of $N$ displacement and $N$ rotational moves for constant $NVT$ simulations, and two additional volume moves for $NPT$ simulations. The amplitude of displacements is chosen such that an acceptance rate of about 40\% is obtained. System sizes are chosen to be similar from one phase to the other, while respecting the crystal symmetry, here, $N_\mathrm{DBI}$=432 and $N_\mathrm{DBN}$=512. For convenience, these simulations report distances in units of the particle diameter, $\sigma=2.54$nm, determined as the shortest center of mass distance between two proteins in contact, and energies in units of $k_\mathrm{B} T_\mathrm{ref}$ with $T_\mathrm{ref}=277$K. 

\subsection*{Estimating the Critical Temperature with Wertheim's Perturbation Theory}
Because Monte Carlo sampling gets increasingly inefficient as temperature is decreases, the critical temperature and the liquid-liquid coexistence regimes are determined using Wertheim's perturbation theory.
The fluid free energy is then approximated as 

\begin{gather}
	\frac{\beta A_f}{N} = \frac{\beta A_\mathrm{HS}}{N} + \frac{\beta A_\mathrm{bond}}{N},
\end{gather}
where $A_\mathrm{HS}$ is the free energy of the hard sphere fluid and $A_\mathrm{bond}$ is the bonding contribution

\begin{gather}
	\frac{\beta A_\mathrm{bond}}{N}=\sum\limits_{a\in\Gamma} \Big(\ln X_a - \frac{X_a}{2}\Big) + \frac{M}{2},
\end{gather}
where the sum runs over all patches, $X_a$ is the bonding probability of patch $a$, and $M$ is the number of patches. Bonding probability is calculated as

\begin{gather}
	X_a = \frac{2}{1+\sqrt{1+4\rho K_a}},
\end{gather}
where $K_a$ quantifies the interaction strength of patch $a$ and is calculated as in Ref.~\citen{Liu:2009},

\begin{gather}
	K_a =4\pi g_\mathrm{ref}^c (e^{\beta\varepsilon_a}-1)\sin^2\Big(\frac{\delta_a}{2}\Big)\sin^2\Big(\frac{\delta_a^\prime}{2}\Big)(\lambda_a-1)\Big(\frac{\Delta\chi_a}{\pi}\Big),
\end{gather}
where $g_\mathrm{ref}^c$ is the contact value of the radial distribution function of a reference fluid, which we take to be the hard sphere fluid for simplicity, and $\varepsilon_a$, $\delta_a$, $\delta_a^\prime$, $\lambda_a$, and $\Delta\chi_a$ are patch parameters. The $(\Delta\chi_a/\pi)$ term takes into account the reduced bonding volume due to the dihedral constraint. The critical temperature, $T_\mathrm{c}$ is then calculated by setting 

\begin{gather}
	\Bigg(\frac{\partial \beta p}{\partial \rho}\Bigg)_{T_\mathrm{c}}=0\\
	\Bigg(\frac{\partial^2 \beta p}{\partial \rho^2}\Bigg)_{T_\mathrm{c}}=0.
\end{gather}
The binodal is then traced using a Maxwell construction at various temperatures. 

Note that for this calculation patches cII, cIV, and cVI, which are collocated, are treated by using $\lambda_{II}$, $\varepsilon=\varepsilon_{II}+\varepsilon_{IV}+\varepsilon_{VI}$, and $\delta_{II}=\delta_{II}^\prime$ as the patch parameters.

\providecommand{\latin}[1]{#1}
\makeatletter
\providecommand{\doi}
  {\begingroup\let\do\@makeother\dospecials
  \catcode`\{=1 \catcode`\}=2 \doi@aux}
\providecommand{\doi@aux}[1]{\endgroup\texttt{#1}}
\makeatother
\providecommand*\mcitethebibliography{\thebibliography}
\csname @ifundefined\endcsname{endmcitethebibliography}
  {\let\endmcitethebibliography\endthebibliography}{}


\begin{mcitethebibliography}{66}
\providecommand*\natexlab[1]{#1}
\providecommand*\mciteSetBstSublistMode[1]{}
\providecommand*\mciteSetBstMaxWidthForm[2]{}
\providecommand*\mciteBstWouldAddEndPuncttrue
  {\def\EndOfBibitem{\unskip.}}
\providecommand*\mciteBstWouldAddEndPunctfalse
  {\let\EndOfBibitem\relax}
\providecommand*\mciteSetBstMidEndSepPunct[3]{}
\providecommand*\mciteSetBstSublistLabelBeginEnd[3]{}
\providecommand*\EndOfBibitem{}
\mciteSetBstSublistMode{f}
\mciteSetBstMaxWidthForm{subitem}{(\alph{mcitesubitemcount})}
\mciteSetBstSublistLabelBeginEnd
  {\mcitemaxwidthsubitemform\space}
  {\relax}
  {\relax}

\bibitem[McManus \latin{et~al.}(2016)McManus, Charbonneau, Zaccarelli, and
  Asherie]{Mcmanus:2016}
McManus,~J.~J.; Charbonneau,~P.; Zaccarelli,~E.; Asherie,~N. The physics of
  protein self-assembly. \emph{Current opinion in colloid \& interface science}
  \textbf{2016}, \emph{22}, 73--79\relax
\mciteBstWouldAddEndPuncttrue
\mciteSetBstMidEndSepPunct{\mcitedefaultmidpunct}
{\mcitedefaultendpunct}{\mcitedefaultseppunct}\relax
\EndOfBibitem
\bibitem[Fusco and Charbonneau(2016)Fusco, and Charbonneau]{Fusco:2016}
Fusco,~D.; Charbonneau,~P. Soft matter perspective on protein crystal assembly.
  \emph{Colloids and Surfaces B: Biointerfaces} \textbf{2016}, \emph{137},
  22--31\relax
\mciteBstWouldAddEndPuncttrue
\mciteSetBstMidEndSepPunct{\mcitedefaultmidpunct}
{\mcitedefaultendpunct}{\mcitedefaultseppunct}\relax
\EndOfBibitem
\bibitem[Hagan(2014)]{Hagan:2014}
Hagan,~M.~F. Modeling viral capsid assembly. \emph{Advances in chemical
  physics} \textbf{2014}, \emph{155}, 1\relax
\mciteBstWouldAddEndPuncttrue
\mciteSetBstMidEndSepPunct{\mcitedefaultmidpunct}
{\mcitedefaultendpunct}{\mcitedefaultseppunct}\relax
\EndOfBibitem
\bibitem[{\v{S}}ari{\'c} \latin{et~al.}(2014){\v{S}}ari{\'c}, Chebaro, Knowles,
  and Frenkel]{Saric:2014}
{\v{S}}ari{\'c},~A.; Chebaro,~Y.~C.; Knowles,~T.~P.; Frenkel,~D. Crucial role
  of nonspecific interactions in amyloid nucleation. \emph{Proceedings of the
  National Academy of Sciences} \textbf{2014}, \emph{111}, 17869--17874\relax
\mciteBstWouldAddEndPuncttrue
\mciteSetBstMidEndSepPunct{\mcitedefaultmidpunct}
{\mcitedefaultendpunct}{\mcitedefaultseppunct}\relax
\EndOfBibitem
\bibitem[Suzuki \latin{et~al.}(2016)Suzuki, Cardone, Restrepo, Zavattieri,
  Baker, and Tezcan]{Suzuki:2016}
Suzuki,~Y.; Cardone,~G.; Restrepo,~D.; Zavattieri,~P.~D.; Baker,~T.~S.;
  Tezcan,~F.~A. Self-assembly of coherently dynamic, auxetic, two-dimensional
  protein crystals. \emph{Nature} \textbf{2016}, \emph{533}, 369\relax
\mciteBstWouldAddEndPuncttrue
\mciteSetBstMidEndSepPunct{\mcitedefaultmidpunct}
{\mcitedefaultendpunct}{\mcitedefaultseppunct}\relax
\EndOfBibitem
\bibitem[Pieters \latin{et~al.}(2016)Pieters, van Eldijk, Nolte, and
  Mecinovi{\'c}]{Pieters:2016}
Pieters,~B.~J.; van Eldijk,~M.~B.; Nolte,~R.~J.; Mecinovi{\'c},~J. Natural
  supramolecular protein assemblies. \emph{Chemical Society Reviews}
  \textbf{2016}, \emph{45}, 24--39\relax
\mciteBstWouldAddEndPuncttrue
\mciteSetBstMidEndSepPunct{\mcitedefaultmidpunct}
{\mcitedefaultendpunct}{\mcitedefaultseppunct}\relax
\EndOfBibitem
\bibitem[Glotzer and Solomon(2007)Glotzer, and Solomon]{Glotzer:2007}
Glotzer,~S.~C.; Solomon,~M.~J. Anisotropy of building blocks and their assembly
  into complex structures. \emph{Nature materials} \textbf{2007}, \emph{6},
  557\relax
\mciteBstWouldAddEndPuncttrue
\mciteSetBstMidEndSepPunct{\mcitedefaultmidpunct}
{\mcitedefaultendpunct}{\mcitedefaultseppunct}\relax
\EndOfBibitem
\bibitem[Huang \latin{et~al.}(2016)Huang, Boyken, and Baker]{Huang:2016}
Huang,~P.-S.; Boyken,~S.~E.; Baker,~D. The coming of age of de novo protein
  design. \emph{Nature} \textbf{2016}, \emph{537}, 320\relax
\mciteBstWouldAddEndPuncttrue
\mciteSetBstMidEndSepPunct{\mcitedefaultmidpunct}
{\mcitedefaultendpunct}{\mcitedefaultseppunct}\relax
\EndOfBibitem
\bibitem[Salgado \latin{et~al.}(2010)Salgado, Radford, and
  Tezcan]{Salgado:2010}
Salgado,~E.~N.; Radford,~R.~J.; Tezcan,~F.~A. Metal-directed protein
  self-assembly. \emph{Accounts of chemical research} \textbf{2010}, \emph{43},
  661--672\relax
\mciteBstWouldAddEndPuncttrue
\mciteSetBstMidEndSepPunct{\mcitedefaultmidpunct}
{\mcitedefaultendpunct}{\mcitedefaultseppunct}\relax
\EndOfBibitem
\bibitem[Koehler~Leman \latin{et~al.}(2015)Koehler~Leman, Ulmschneider, and
  Gray]{Koehler:2015}
Koehler~Leman,~J.; Ulmschneider,~M.~B.; Gray,~J.~J. Computational modeling of
  membrane proteins. \emph{Proteins: Structure, Function, and Bioinformatics}
  \textbf{2015}, \emph{83}, 1--24\relax
\mciteBstWouldAddEndPuncttrue
\mciteSetBstMidEndSepPunct{\mcitedefaultmidpunct}
{\mcitedefaultendpunct}{\mcitedefaultseppunct}\relax
\EndOfBibitem
\bibitem[Altan and Charbonneau(2019)Altan, and Charbonneau]{Altan:2019}
Altan,~I.; Charbonneau,~P. \emph{Protein Self-Assembly}; Springer, 2019; pp
  209--227\relax
\mciteBstWouldAddEndPuncttrue
\mciteSetBstMidEndSepPunct{\mcitedefaultmidpunct}
{\mcitedefaultendpunct}{\mcitedefaultseppunct}\relax
\EndOfBibitem
\bibitem[Vega \latin{et~al.}(2008)Vega, Sanz, Abascal, and Noya]{Vega2008}
Vega,~C.; Sanz,~E.; Abascal,~J.; Noya,~E. Determination of phase diagrams via
  computer simulation: methodology and applications to water, electrolytes and
  proteins. \emph{Journal of Physics: Condensed Matter} \textbf{2008},
  \emph{20}, 153101\relax
\mciteBstWouldAddEndPuncttrue
\mciteSetBstMidEndSepPunct{\mcitedefaultmidpunct}
{\mcitedefaultendpunct}{\mcitedefaultseppunct}\relax
\EndOfBibitem
\bibitem[Romano \latin{et~al.}(2010)Romano, Sanz, and Sciortino]{Romano:2010}
Romano,~F.; Sanz,~E.; Sciortino,~F. Phase diagram of a tetrahedral patchy
  particle model for different interaction ranges. \emph{The Journal of
  Chemical Physics} \textbf{2010}, \emph{132}, 184501\relax
\mciteBstWouldAddEndPuncttrue
\mciteSetBstMidEndSepPunct{\mcitedefaultmidpunct}
{\mcitedefaultendpunct}{\mcitedefaultseppunct}\relax
\EndOfBibitem
\bibitem[ten Wolde and Frenkel(1997)ten Wolde, and Frenkel]{tenWolde:1997}
ten Wolde,~P.~R.; Frenkel,~D. Enhancement of protein crystal nucleation by
  critical density fluctuations. \emph{Science} \textbf{1997}, \emph{277},
  1975--1978\relax
\mciteBstWouldAddEndPuncttrue
\mciteSetBstMidEndSepPunct{\mcitedefaultmidpunct}
{\mcitedefaultendpunct}{\mcitedefaultseppunct}\relax
\EndOfBibitem
\bibitem[Romano \latin{et~al.}(2009)Romano, Sanz, and Sciortino]{Romano:2009}
Romano,~F.; Sanz,~E.; Sciortino,~F. Role of the range in the fluid- crystal
  coexistence for a patchy particle model. \emph{The Journal of Physical
  Chemistry B} \textbf{2009}, \emph{113}, 15133--15136\relax
\mciteBstWouldAddEndPuncttrue
\mciteSetBstMidEndSepPunct{\mcitedefaultmidpunct}
{\mcitedefaultendpunct}{\mcitedefaultseppunct}\relax
\EndOfBibitem
\bibitem[Lomakin \latin{et~al.}(1999)Lomakin, Asherie, and
  Benedek]{Lomakin:1999}
Lomakin,~A.; Asherie,~N.; Benedek,~G.~B. Aeolotopic interactions of globular
  proteins. \emph{Proceedings of the National Academy of Sciences}
  \textbf{1999}, \emph{96}, 9465--9468\relax
\mciteBstWouldAddEndPuncttrue
\mciteSetBstMidEndSepPunct{\mcitedefaultmidpunct}
{\mcitedefaultendpunct}{\mcitedefaultseppunct}\relax
\EndOfBibitem
\bibitem[Bianchi \latin{et~al.}(2011)Bianchi, Blaak, and Likos]{Bianchi:2011}
Bianchi,~E.; Blaak,~R.; Likos,~C.~N. Patchy colloids: state of the art and
  perspectives. \emph{Physical Chemistry Chemical Physics} \textbf{2011},
  \emph{13}, 6397--6410\relax
\mciteBstWouldAddEndPuncttrue
\mciteSetBstMidEndSepPunct{\mcitedefaultmidpunct}
{\mcitedefaultendpunct}{\mcitedefaultseppunct}\relax
\EndOfBibitem
\bibitem[McPherson(2004)]{McPherson:2004}
McPherson,~A. Introduction to protein crystallization. \emph{Methods}
  \textbf{2004}, \emph{34}, 254--265\relax
\mciteBstWouldAddEndPuncttrue
\mciteSetBstMidEndSepPunct{\mcitedefaultmidpunct}
{\mcitedefaultendpunct}{\mcitedefaultseppunct}\relax
\EndOfBibitem
\bibitem[Pande \latin{et~al.}(2009)Pande, Zhang, Banerjee, Puttamadappa,
  Shekhtman, and Pande]{Pande:2009}
Pande,~A.; Zhang,~J.; Banerjee,~P.~R.; Puttamadappa,~S.~S.; Shekhtman,~A.;
  Pande,~J. NMR study of the cataract-linked P23T mutant of human
  $\gamma$D-crystallin shows minor changes in hydrophobic patches that reflect
  its retrograde solubility. \emph{Biochemical and biophysical research
  communications} \textbf{2009}, \emph{382}, 196--199\relax
\mciteBstWouldAddEndPuncttrue
\mciteSetBstMidEndSepPunct{\mcitedefaultmidpunct}
{\mcitedefaultendpunct}{\mcitedefaultseppunct}\relax
\EndOfBibitem
\bibitem[Pande \latin{et~al.}(2010)Pande, Ghosh, Banerjee, and
  Pande]{Pande:2010}
Pande,~A.; Ghosh,~K.~S.; Banerjee,~P.~R.; Pande,~J. Increase in surface
  hydrophobicity of the cataract-associated P23T mutant of human
  $\gamma$D-crystallin is responsible for its dramatically lower, retrograde
  solubility. \emph{Biochemistry} \textbf{2010}, \emph{49}, 6122--6129\relax
\mciteBstWouldAddEndPuncttrue
\mciteSetBstMidEndSepPunct{\mcitedefaultmidpunct}
{\mcitedefaultendpunct}{\mcitedefaultseppunct}\relax
\EndOfBibitem
\bibitem[Vekilov \latin{et~al.}(2002)Vekilov, Feeling-Taylor, Yau, and
  Petsev]{Vekilov:2002}
Vekilov,~P.~G.; Feeling-Taylor,~A.; Yau,~S.-T.; Petsev,~D. Solvent entropy
  contribution to the free energy of protein crystallization. \emph{Acta
  Crystallographica Section D: Biological Crystallography} \textbf{2002},
  \emph{58}, 1611--1616\relax
\mciteBstWouldAddEndPuncttrue
\mciteSetBstMidEndSepPunct{\mcitedefaultmidpunct}
{\mcitedefaultendpunct}{\mcitedefaultseppunct}\relax
\EndOfBibitem
\bibitem[Petsev \latin{et~al.}(2001)Petsev, Thomas, Yau, Tsekova, Nanev,
  Wilson, and Vekilov]{Petsev:2001}
Petsev,~D.~N.; Thomas,~B.~R.; Yau,~S.-T.; Tsekova,~D.; Nanev,~C.;
  Wilson,~W.~W.; Vekilov,~P.~G. Temperature-independent solubility and
  interactions between apoferritin monomers and dimers in solution.
  \emph{Journal of crystal growth} \textbf{2001}, \emph{232}, 21--29\relax
\mciteBstWouldAddEndPuncttrue
\mciteSetBstMidEndSepPunct{\mcitedefaultmidpunct}
{\mcitedefaultendpunct}{\mcitedefaultseppunct}\relax
\EndOfBibitem
\bibitem[Shiryayev \latin{et~al.}(2005)Shiryayev, Pagan, Gunton, Rhen, Saxena,
  and Lookman]{Shiryayev:2005}
Shiryayev,~A.; Pagan,~D.~L.; Gunton,~J.~D.; Rhen,~D.; Saxena,~A.; Lookman,~T.
  Role of solvent for globular proteins in solution. \emph{The Journal of
  chemical physics} \textbf{2005}, \emph{122}, 234911\relax
\mciteBstWouldAddEndPuncttrue
\mciteSetBstMidEndSepPunct{\mcitedefaultmidpunct}
{\mcitedefaultendpunct}{\mcitedefaultseppunct}\relax
\EndOfBibitem
\bibitem[Wentzel and Gunton(2008)Wentzel, and Gunton]{Wentzel:2008}
Wentzel,~N.; Gunton,~J.~D. Effect of solvent on the phase diagram of a simple
  anisotropic model of globular proteins. \emph{The Journal of Physical
  Chemistry B} \textbf{2008}, \emph{112}, 7803--7809\relax
\mciteBstWouldAddEndPuncttrue
\mciteSetBstMidEndSepPunct{\mcitedefaultmidpunct}
{\mcitedefaultendpunct}{\mcitedefaultseppunct}\relax
\EndOfBibitem
\bibitem[Moelbert and De~Los~Rios(2003)Moelbert, and
  De~Los~Rios]{Moelbert:2003}
Moelbert,~S.; De~Los~Rios,~P. Hydrophobic interaction model for upper and lower
  critical solution temperatures. \emph{Macromolecules} \textbf{2003},
  \emph{36}, 5845--5853\relax
\mciteBstWouldAddEndPuncttrue
\mciteSetBstMidEndSepPunct{\mcitedefaultmidpunct}
{\mcitedefaultendpunct}{\mcitedefaultseppunct}\relax
\EndOfBibitem
\bibitem[Lee and Graziano(1996)Lee, and Graziano]{Lee:1996}
Lee,~B.; Graziano,~G. A two-state model of hydrophobic hydration that produces
  compensating enthalpy and entropy changes. \emph{Journal of the American
  Chemical Society} \textbf{1996}, \emph{118}, 5163--5168\relax
\mciteBstWouldAddEndPuncttrue
\mciteSetBstMidEndSepPunct{\mcitedefaultmidpunct}
{\mcitedefaultendpunct}{\mcitedefaultseppunct}\relax
\EndOfBibitem
\bibitem[Muller(1990)]{Muller:1990}
Muller,~N. Search for a realistic view of hydrophobic effects. \emph{Accounts
  of Chemical Research} \textbf{1990}, \emph{23}, 23--28\relax
\mciteBstWouldAddEndPuncttrue
\mciteSetBstMidEndSepPunct{\mcitedefaultmidpunct}
{\mcitedefaultendpunct}{\mcitedefaultseppunct}\relax
\EndOfBibitem
\bibitem[Berman \latin{et~al.}(2000)Berman, Westbrook, Feng, Gilliland, Bhat,
  Weissig, Shindyalov, and Bourne]{Berman:2000}
Berman,~H.~M.; Westbrook,~J.; Feng,~Z.; Gilliland,~G.; Bhat,~T.~N.;
  Weissig,~H.; Shindyalov,~I.~N.; Bourne,~P.~E. The protein data bank.
  \emph{Nucleic acids research} \textbf{2000}, \emph{28}, 235--242\relax
\mciteBstWouldAddEndPuncttrue
\mciteSetBstMidEndSepPunct{\mcitedefaultmidpunct}
{\mcitedefaultendpunct}{\mcitedefaultseppunct}\relax
\EndOfBibitem
\bibitem[James \latin{et~al.}(2015)James, Quinn, and McManus]{James:2015}
James,~S.; Quinn,~M.~K.; McManus,~J.~J. The self assembly of proteins; probing
  patchy protein interactions. \emph{Physical Chemistry Chemical Physics}
  \textbf{2015}, \emph{17}, 5413--5420\relax
\mciteBstWouldAddEndPuncttrue
\mciteSetBstMidEndSepPunct{\mcitedefaultmidpunct}
{\mcitedefaultendpunct}{\mcitedefaultseppunct}\relax
\EndOfBibitem
\bibitem[Khan \latin{et~al.}(2019)Khan, James, Quinn, Altan, Charbonneau, and
  McManus]{Khan:2019}
Khan,~A.~R.; James,~S.; Quinn,~M.~K.; Altan,~I.; Charbonneau,~P.;
  McManus,~J.~J. Temperature-dependent interactions explain normal and inverted
  solubility in a $\gamma$D-crystallin mutant. \emph{Biophysical Journal}
  \textbf{2019}, \relax
\mciteBstWouldAddEndPunctfalse
\mciteSetBstMidEndSepPunct{\mcitedefaultmidpunct}
{}{\mcitedefaultseppunct}\relax
\EndOfBibitem
\bibitem[McManus \latin{et~al.}(2007)McManus, Lomakin, Ogun, Pande, Basan,
  Pande, and Benedek]{Mcmanus:2007}
McManus,~J.~J.; Lomakin,~A.; Ogun,~O.; Pande,~A.; Basan,~M.; Pande,~J.;
  Benedek,~G.~B. Altered phase diagram due to a single point mutation in human
  $\gamma$D-crystallin. \emph{Proceedings of the National Academy of Sciences}
  \textbf{2007}, \emph{104}, 16856--16861\relax
\mciteBstWouldAddEndPuncttrue
\mciteSetBstMidEndSepPunct{\mcitedefaultmidpunct}
{\mcitedefaultendpunct}{\mcitedefaultseppunct}\relax
\EndOfBibitem
\bibitem[{Schr\"odinger, LLC}(2015)]{PyMol}
{Schr\"odinger, LLC}, \relax
\mciteBstWouldAddEndPunctfalse
\mciteSetBstMidEndSepPunct{\mcitedefaultmidpunct}
{}{\mcitedefaultseppunct}\relax
\EndOfBibitem
\bibitem[Fusco \latin{et~al.}(2014)Fusco, Barnum, Bruno, Luft, Snell,
  Mukherjee, and Charbonneau]{Fusco:2014b}
Fusco,~D.; Barnum,~T.~J.; Bruno,~A.~E.; Luft,~J.~R.; Snell,~E.~H.;
  Mukherjee,~S.; Charbonneau,~P. Statistical analysis of crystallization
  database links protein physico-chemical features with crystallization
  mechanisms. \emph{PLoS One} \textbf{2014}, \emph{9}, e101123\relax
\mciteBstWouldAddEndPuncttrue
\mciteSetBstMidEndSepPunct{\mcitedefaultmidpunct}
{\mcitedefaultendpunct}{\mcitedefaultseppunct}\relax
\EndOfBibitem
\bibitem[Kyte and Doolittle(1982)Kyte, and Doolittle]{Kyte:1982}
Kyte,~J.; Doolittle,~R.~F. A simple method for displaying the hydropathic
  character of a protein. \emph{Journal of molecular biology} \textbf{1982},
  \emph{157}, 105--132\relax
\mciteBstWouldAddEndPuncttrue
\mciteSetBstMidEndSepPunct{\mcitedefaultmidpunct}
{\mcitedefaultendpunct}{\mcitedefaultseppunct}\relax
\EndOfBibitem
\bibitem[Wimley and White(1996)Wimley, and White]{Wimley:1996}
Wimley,~W.~C.; White,~S.~H. Experimentally determined hydrophobicity scale for
  proteins at membrane interfaces. \emph{Nature Structural and Molecular
  Biology} \textbf{1996}, \emph{3}, 842\relax
\mciteBstWouldAddEndPuncttrue
\mciteSetBstMidEndSepPunct{\mcitedefaultmidpunct}
{\mcitedefaultendpunct}{\mcitedefaultseppunct}\relax
\EndOfBibitem
\bibitem[Hessa \latin{et~al.}(2005)Hessa, Kim, Bihlmaier, Lundin, Boekel,
  Andersson, Nilsson, White, and von Heijne]{Hessa:2005}
Hessa,~T.; Kim,~H.; Bihlmaier,~K.; Lundin,~C.; Boekel,~J.; Andersson,~H.;
  Nilsson,~I.; White,~S.~H.; von Heijne,~G. Recognition of transmembrane
  helices by the endoplasmic reticulum translocon. \emph{Nature} \textbf{2005},
  \emph{433}, 377\relax
\mciteBstWouldAddEndPuncttrue
\mciteSetBstMidEndSepPunct{\mcitedefaultmidpunct}
{\mcitedefaultendpunct}{\mcitedefaultseppunct}\relax
\EndOfBibitem
\bibitem[Moon and Fleming(2011)Moon, and Fleming]{Moon:2011}
Moon,~C.~P.; Fleming,~K.~G. Side-chain hydrophobicity scale derived from
  transmembrane protein folding into lipid bilayers. \emph{Proceedings of the
  National Academy of Sciences} \textbf{2011}, \emph{108}, 10174--10177\relax
\mciteBstWouldAddEndPuncttrue
\mciteSetBstMidEndSepPunct{\mcitedefaultmidpunct}
{\mcitedefaultendpunct}{\mcitedefaultseppunct}\relax
\EndOfBibitem
\bibitem[Zhao and London(2006)Zhao, and London]{Zhao:2006}
Zhao,~G.; London,~E. An amino acid “transmembrane tendency” scale that
  approaches the theoretical limit to accuracy for prediction of transmembrane
  helices: relationship to biological hydrophobicity. \emph{Protein science}
  \textbf{2006}, \emph{15}, 1987--2001\relax
\mciteBstWouldAddEndPuncttrue
\mciteSetBstMidEndSepPunct{\mcitedefaultmidpunct}
{\mcitedefaultendpunct}{\mcitedefaultseppunct}\relax
\EndOfBibitem
\bibitem[Basak \latin{et~al.}(2003)Basak, Bateman, Slingsby, Pande, Asherie,
  Ogun, Benedek, and Pande]{Basak:2003}
Basak,~A.; Bateman,~O.; Slingsby,~C.; Pande,~A.; Asherie,~N.; Ogun,~O.;
  Benedek,~G.~B.; Pande,~J. High-resolution X-ray crystal structures of human
  $\gamma$D crystallin (1.25 {\AA}) and the R58H mutant (1.15 {\AA}) associated
  with aculeiform cataract. \emph{Journal of molecular biology} \textbf{2003},
  \emph{328}, 1137--1147\relax
\mciteBstWouldAddEndPuncttrue
\mciteSetBstMidEndSepPunct{\mcitedefaultmidpunct}
{\mcitedefaultendpunct}{\mcitedefaultseppunct}\relax
\EndOfBibitem
\bibitem[Ji \latin{et~al.}(2013)Ji, Koharudin, Jung, and Gronenborn]{Ji:2013}
Ji,~F.; Koharudin,~L.~M.; Jung,~J.; Gronenborn,~A.~M. Crystal structure of the
  cataract-causing P23T $\gamma$D-crystallin mutant. \emph{Proteins: Structure,
  Function, and Bioinformatics} \textbf{2013}, \emph{81}, 1493--1498\relax
\mciteBstWouldAddEndPuncttrue
\mciteSetBstMidEndSepPunct{\mcitedefaultmidpunct}
{\mcitedefaultendpunct}{\mcitedefaultseppunct}\relax
\EndOfBibitem
\bibitem[Kmoch \latin{et~al.}(2000)Kmoch, Brynda, Asfaw, Bezou{\v{s}}ka,
  Nov{\'a}k, {\v{R}}ez{\'a}{\v{c}}ov{\'a}, Ondrov{\'a}, Filipec,
  Sedl{\'a}{\v{c}}ek, and Elleder]{Kmoch:2000}
Kmoch,~S.; Brynda,~J.; Asfaw,~B.; Bezou{\v{s}}ka,~K.; Nov{\'a}k,~P.;
  {\v{R}}ez{\'a}{\v{c}}ov{\'a},~P.; Ondrov{\'a},~L.; Filipec,~M.;
  Sedl{\'a}{\v{c}}ek,~J.; Elleder,~M. Link between a novel human
  $\gamma$D-crystallin allele and a unique cataract phenotype explained by
  protein crystallography. \emph{Human molecular genetics} \textbf{2000},
  \emph{9}, 1779--1786\relax
\mciteBstWouldAddEndPuncttrue
\mciteSetBstMidEndSepPunct{\mcitedefaultmidpunct}
{\mcitedefaultendpunct}{\mcitedefaultseppunct}\relax
\EndOfBibitem
\bibitem[{\v{S}}ali and Blundell(1993){\v{S}}ali, and Blundell]{sali:1993}
{\v{S}}ali,~A.; Blundell,~T.~L. Comparative protein modelling by satisfaction
  of spatial restraints. \emph{Journal of molecular biology} \textbf{1993},
  \emph{234}, 779--815\relax
\mciteBstWouldAddEndPuncttrue
\mciteSetBstMidEndSepPunct{\mcitedefaultmidpunct}
{\mcitedefaultendpunct}{\mcitedefaultseppunct}\relax
\EndOfBibitem
\bibitem[Pettersen \latin{et~al.}(2004)Pettersen, Goddard, Huang, Couch,
  Greenblatt, Meng, and Ferrin]{Pettersen:2004}
Pettersen,~E.~F.; Goddard,~T.~D.; Huang,~C.~C.; Couch,~G.~S.;
  Greenblatt,~D.~M.; Meng,~E.~C.; Ferrin,~T.~E. UCSF Chimera—a visualization
  system for exploratory research and analysis. \emph{Journal of computational
  chemistry} \textbf{2004}, \emph{25}, 1605--1612\relax
\mciteBstWouldAddEndPuncttrue
\mciteSetBstMidEndSepPunct{\mcitedefaultmidpunct}
{\mcitedefaultendpunct}{\mcitedefaultseppunct}\relax
\EndOfBibitem
\bibitem[Fusco \latin{et~al.}(2014)Fusco, Headd, De~Simone, Wang, and
  Charbonneau]{Fusco:2014}
Fusco,~D.; Headd,~J.~J.; De~Simone,~A.; Wang,~J.; Charbonneau,~P.
  Characterizing protein crystal contacts and their role in crystallization:
  rubredoxin as a case study. \emph{Soft matter} \textbf{2014}, \emph{10},
  290--302\relax
\mciteBstWouldAddEndPuncttrue
\mciteSetBstMidEndSepPunct{\mcitedefaultmidpunct}
{\mcitedefaultendpunct}{\mcitedefaultseppunct}\relax
\EndOfBibitem
\bibitem[Berendsen \latin{et~al.}(1995)Berendsen, van~der Spoel, and van
  Drunen]{Berendsen:1995}
Berendsen,~H.~J.; van~der Spoel,~D.; van Drunen,~R. GROMACS: a message-passing
  parallel molecular dynamics implementation. \emph{Computer physics
  communications} \textbf{1995}, \emph{91}, 43--56\relax
\mciteBstWouldAddEndPuncttrue
\mciteSetBstMidEndSepPunct{\mcitedefaultmidpunct}
{\mcitedefaultendpunct}{\mcitedefaultseppunct}\relax
\EndOfBibitem
\bibitem[K{\"a}stner(2011)]{Kastner:2011}
K{\"a}stner,~J. Umbrella sampling. \emph{Wiley Interdisciplinary Reviews:
  Computational Molecular Science} \textbf{2011}, \emph{1}, 932--942\relax
\mciteBstWouldAddEndPuncttrue
\mciteSetBstMidEndSepPunct{\mcitedefaultmidpunct}
{\mcitedefaultendpunct}{\mcitedefaultseppunct}\relax
\EndOfBibitem
\bibitem[Wickstrom \latin{et~al.}(2009)Wickstrom, Okur, and
  Simmerling]{Wickstrom:2009}
Wickstrom,~L.; Okur,~A.; Simmerling,~C. Evaluating the performance of the
  ff99SB force field based on NMR scalar coupling data. \emph{Biophysical
  journal} \textbf{2009}, \emph{97}, 853--856\relax
\mciteBstWouldAddEndPuncttrue
\mciteSetBstMidEndSepPunct{\mcitedefaultmidpunct}
{\mcitedefaultendpunct}{\mcitedefaultseppunct}\relax
\EndOfBibitem
\bibitem[Altan \latin{et~al.}(2018)Altan, Fusco, Afonine, and
  Charbonneau]{Altan:2018}
Altan,~I.; Fusco,~D.; Afonine,~P.~V.; Charbonneau,~P. Learning about
  Biomolecular Solvation from Water in Protein Crystals. \emph{The Journal of
  Physical Chemistry B} \textbf{2018}, \emph{122}, 2475--2486\relax
\mciteBstWouldAddEndPuncttrue
\mciteSetBstMidEndSepPunct{\mcitedefaultmidpunct}
{\mcitedefaultendpunct}{\mcitedefaultseppunct}\relax
\EndOfBibitem
\bibitem[Henzler-Wildman and Kern(2007)Henzler-Wildman, and Kern]{Henzler:2007}
Henzler-Wildman,~K.; Kern,~D. Dynamic personalities of proteins. \emph{Nature}
  \textbf{2007}, \emph{450}, 964\relax
\mciteBstWouldAddEndPuncttrue
\mciteSetBstMidEndSepPunct{\mcitedefaultmidpunct}
{\mcitedefaultendpunct}{\mcitedefaultseppunct}\relax
\EndOfBibitem
\bibitem[Silverstein \latin{et~al.}(1999)Silverstein, Haymet, and
  Dill]{Silverstein:1999}
Silverstein,~K.~A.; Haymet,~A.; Dill,~K.~A. Molecular model of hydrophobic
  solvation. \emph{The Journal of chemical physics} \textbf{1999}, \emph{111},
  8000--8009\relax
\mciteBstWouldAddEndPuncttrue
\mciteSetBstMidEndSepPunct{\mcitedefaultmidpunct}
{\mcitedefaultendpunct}{\mcitedefaultseppunct}\relax
\EndOfBibitem
\bibitem[Silverstein \latin{et~al.}(1998)Silverstein, Haymet, and
  Dill]{Silverstein:1998}
Silverstein,~K.~A.; Haymet,~A.; Dill,~K.~A. A simple model of water and the
  hydrophobic effect. \emph{Journal of the American Chemical Society}
  \textbf{1998}, \emph{120}, 3166--3175\relax
\mciteBstWouldAddEndPuncttrue
\mciteSetBstMidEndSepPunct{\mcitedefaultmidpunct}
{\mcitedefaultendpunct}{\mcitedefaultseppunct}\relax
\EndOfBibitem
\bibitem[Feyereisen \latin{et~al.}(1996)Feyereisen, Feller, and
  Dixon]{Feyereisen:1996}
Feyereisen,~M.~W.; Feller,~D.; Dixon,~D.~A. Hydrogen bond energy of the water
  dimer. \emph{The Journal of Physical Chemistry} \textbf{1996}, \emph{100},
  2993--2997\relax
\mciteBstWouldAddEndPuncttrue
\mciteSetBstMidEndSepPunct{\mcitedefaultmidpunct}
{\mcitedefaultendpunct}{\mcitedefaultseppunct}\relax
\EndOfBibitem
\bibitem[Wertheim(1984)]{Wertheim:1984}
Wertheim,~M. Fluids with highly directional attractive forces. I. Statistical
  thermodynamics. \emph{Journal of statistical physics} \textbf{1984},
  \emph{35}, 19--34\relax
\mciteBstWouldAddEndPuncttrue
\mciteSetBstMidEndSepPunct{\mcitedefaultmidpunct}
{\mcitedefaultendpunct}{\mcitedefaultseppunct}\relax
\EndOfBibitem
\bibitem[Wertheim(1984)]{Wertheim:1984b}
Wertheim,~M. Fluids with highly directional attractive forces. II.
  Thermodynamic perturbation theory and integral equations. \emph{Journal of
  statistical physics} \textbf{1984}, \emph{35}, 35--47\relax
\mciteBstWouldAddEndPuncttrue
\mciteSetBstMidEndSepPunct{\mcitedefaultmidpunct}
{\mcitedefaultendpunct}{\mcitedefaultseppunct}\relax
\EndOfBibitem
\bibitem[Sear(1999)]{Sear:1999}
Sear,~R.~P. Phase behavior of a simple model of globular proteins. \emph{The
  Journal of chemical physics} \textbf{1999}, \emph{111}, 4800--4806\relax
\mciteBstWouldAddEndPuncttrue
\mciteSetBstMidEndSepPunct{\mcitedefaultmidpunct}
{\mcitedefaultendpunct}{\mcitedefaultseppunct}\relax
\EndOfBibitem
\bibitem[de~las Heras and da~Gama(2016)de~las Heras, and da~Gama]{de:2016}
de~las Heras,~D.; da~Gama,~M. M.~T. Temperature (de) activated patchy colloidal
  particles. \emph{Journal of Physics: Condensed Matter} \textbf{2016},
  \emph{28}, 244008\relax
\mciteBstWouldAddEndPuncttrue
\mciteSetBstMidEndSepPunct{\mcitedefaultmidpunct}
{\mcitedefaultendpunct}{\mcitedefaultseppunct}\relax
\EndOfBibitem
\bibitem[Geerts and Eiser(2010)Geerts, and Eiser]{Geerts:2010}
Geerts,~N.; Eiser,~E. DNA-functionalized colloids: Physical properties and
  applications. \emph{Soft Matter} \textbf{2010}, \emph{6}, 4647--4660\relax
\mciteBstWouldAddEndPuncttrue
\mciteSetBstMidEndSepPunct{\mcitedefaultmidpunct}
{\mcitedefaultendpunct}{\mcitedefaultseppunct}\relax
\EndOfBibitem
\bibitem[Chen and Siepmann(2000)Chen, and Siepmann]{Chen:2000}
Chen,~B.; Siepmann,~J.~I. A novel Monte Carlo algorithm for simulating strongly
  associating fluids: Applications to water, hydrogen fluoride, and acetic
  acid. \emph{The Journal of Physical Chemistry B} \textbf{2000}, \emph{104},
  8725--8734\relax
\mciteBstWouldAddEndPuncttrue
\mciteSetBstMidEndSepPunct{\mcitedefaultmidpunct}
{\mcitedefaultendpunct}{\mcitedefaultseppunct}\relax
\EndOfBibitem
\bibitem[Bernard \latin{et~al.}(2009)Bernard, Krauth, and Wilson]{Bernard:2009}
Bernard,~E.~P.; Krauth,~W.; Wilson,~D.~B. Event-chain Monte Carlo algorithms
  for hard-sphere systems. \emph{Physical Review E} \textbf{2009}, \emph{80},
  056704\relax
\mciteBstWouldAddEndPuncttrue
\mciteSetBstMidEndSepPunct{\mcitedefaultmidpunct}
{\mcitedefaultendpunct}{\mcitedefaultseppunct}\relax
\EndOfBibitem
\bibitem[Kern and Frenkel(2003)Kern, and Frenkel]{Kern:2003}
Kern,~N.; Frenkel,~D. Fluid--fluid coexistence in colloidal systems with
  short-ranged strongly directional attraction. \emph{The Journal of chemical
  physics} \textbf{2003}, \emph{118}, 9882--9889\relax
\mciteBstWouldAddEndPuncttrue
\mciteSetBstMidEndSepPunct{\mcitedefaultmidpunct}
{\mcitedefaultendpunct}{\mcitedefaultseppunct}\relax
\EndOfBibitem
\bibitem[Fusco and Charbonneau(2013)Fusco, and Charbonneau]{Fusco:2013}
Fusco,~D.; Charbonneau,~P. Crystallization of asymmetric patchy models for
  globular proteins in solution. \emph{Physical Review E} \textbf{2013},
  \emph{88}, 012721\relax
\mciteBstWouldAddEndPuncttrue
\mciteSetBstMidEndSepPunct{\mcitedefaultmidpunct}
{\mcitedefaultendpunct}{\mcitedefaultseppunct}\relax
\EndOfBibitem
\bibitem[Frenkel and Ladd(1984)Frenkel, and Ladd]{Frenkel:1984}
Frenkel,~D.; Ladd,~A.~J. New Monte Carlo method to compute the free energy of
  arbitrary solids. Application to the fcc and hcp phases of hard spheres.
  \emph{The Journal of chemical physics} \textbf{1984}, \emph{81},
  3188--3193\relax
\mciteBstWouldAddEndPuncttrue
\mciteSetBstMidEndSepPunct{\mcitedefaultmidpunct}
{\mcitedefaultendpunct}{\mcitedefaultseppunct}\relax
\EndOfBibitem
\bibitem[Liu \latin{et~al.}(2009)Liu, Kumar, Sciortino, and Evans]{Liu:2009}
Liu,~H.; Kumar,~S.~K.; Sciortino,~F.; Evans,~G.~T. Vapor-liquid coexistence of
  fluids with attractive patches: An application of Wertheim’s theory of
  association. \emph{The Journal of chemical physics} \textbf{2009},
  \emph{130}, 044902\relax
\mciteBstWouldAddEndPuncttrue
\mciteSetBstMidEndSepPunct{\mcitedefaultmidpunct}
{\mcitedefaultendpunct}{\mcitedefaultseppunct}\relax
\EndOfBibitem
\bibitem[Kessler \latin{et~al.}(2015)Kessler, Elgabarty, Spura, Karhan,
  Partovi-Azar, Hassanali, and Kuhne]{Kessler:2015}
Kessler,~J.; Elgabarty,~H.; Spura,~T.; Karhan,~K.; Partovi-Azar,~P.;
  Hassanali,~A.~A.; Kuhne,~T.~D. Structure and dynamics of the instantaneous
  water/vapor interface revisited by path-integral and ab initio molecular
  dynamics simulations. \emph{The Journal of Physical Chemistry B}
  \textbf{2015}, \emph{119}, 10079--10086\relax
\mciteBstWouldAddEndPuncttrue
\mciteSetBstMidEndSepPunct{\mcitedefaultmidpunct}
{\mcitedefaultendpunct}{\mcitedefaultseppunct}\relax
\EndOfBibitem
\bibitem[Tainter \latin{et~al.}(2015)Tainter, Shi, and Skinner]{Tainter:2015}
Tainter,~C.~J.; Shi,~L.; Skinner,~J.~L. Reparametrized E3B (explicit
  three-body) water model using the TIP4P/2005 model as a reference.
  \emph{Journal of chemical theory and computation} \textbf{2015}, \emph{11},
  2268--2277\relax
\mciteBstWouldAddEndPuncttrue
\mciteSetBstMidEndSepPunct{\mcitedefaultmidpunct}
{\mcitedefaultendpunct}{\mcitedefaultseppunct}\relax
\EndOfBibitem
\end{mcitethebibliography}
\end{document}